# Multimodal Knowledge Aggregation for Microbial Keratitis: Geographic Confounding and Shortcut Risk in Pooled Multicountry Data

Yiqing Wang, MS[1]; Maria A. Woodward, MD, MSc[2,3]; Ziyun Yang, PhD[1]; N. Venkatesh Prajna, DO, DNB[4]; Chunming He, ME[1]; Leslie M. Niziol, MS[2]; Mercy Pawar, MS[2]; Ming-Chen Lu, MS[2]; Guillermo Amescua, MD[5]; Rachel A. F. Wozniak, MD, PhD[6]; Sejal Amin, MD[7]; Abinaya Krishnan, MS[4]; Prabhleen Kochar, MS[4]; Sina Farsiu, PhD[1,8]

Affiliations

1 Department of Biomedical Engineering, Duke University, Durham, NC, USA.

2 Department of Ophthalmology and Visual Sciences, University of Michigan, Ann Arbor, MI, USA.

3 Institute for Healthcare Policy and Innovation, University of Michigan, Ann Arbor, MI, USA.

4 Department of Cornea and Refractive Surgery Services, Aravind Eye Care System, Madurai, Tamil Nadu, India.

5 Bascom Palmer Eye Institute, Department of Ophthalmology, University of Miami Miller School of Medicine, Miami, FL, USA.

6 Flaum Eye Institute, Department of Ophthalmology, University of Rochester Medical Center, Rochester, NY, USA.

7 Department of Ophthalmology, Henry Ford Hospital, Detroit, MI, USA.

8 Duke Eye Center, Duke University School of Medicine, Durham, NC, USA.

Corresponding Author and Address for Reprints

Sina Farsiu, PhD; Department of Biomedical Engineering, Duke University; Duke Eye Center
2351 Erwin Road, DUMC 3802, 103 Wadsworth Building, Durham, NC 27710, USA
Telephone: +1 919-684-6642; Email: sina.farsiu@duke.edu

Running Head: Multimodal Keratitis Classification

Keywords: microbial keratitis; multimodal artificial intelligence; slit-lamp photography; geographic confounding; shortcut learning

Online-only material: Appendix 1, Tables 7–13, and Figure 4 are available at https://www.aaojournal.org.

Meeting Presentation: This work has not been presented previously at a scientific meeting.

Financial Support: This work was supported in part by the National Institutes of Health, Bethesda, Maryland (R01EY036418-01 [M.A.W. and S.F.], P30 EY005722 [S.F.], and 2P30EY007003-36 [M.A.W. and M.C.L.]); a Research to Prevent Blindness Career Advancement Award, New York, New York (M.A.W.); and an unrestricted grant from Research to Prevent Blindness, New York, New York, to Duke University. The sponsor or funding organizations had no role in the design or conduct of the research; collection, management, analysis, or interpretation of the data; preparation, review, or approval of the manuscript; or the decision to submit the manuscript for publication.

Conflict of Interest: No conflicting relationship exists for any author.

Abbreviations and Acronyms: AI, artificial intelligence; AQUA, Automated Quantitative Ulcer Analysis; AUROC, area under the receiver operating characteristic curve; B, diffuse blue-light illumination; CI, confidence interval; CMCL, cross-modality contrastive learning; Grad-CAM, gradient-weighted class activation mapping; Group DRO, group distributionally robust optimization; IRB, institutional review board; MK, microbial keratitis; MLP, multilayer perceptron; PCR, polymerase chain reaction; PVT, pyramid vision transformer; ROC, receiver operating characteristic; ScS, sclerotic-scatter illumination; SD, standard deviation; SLP, slit-lamp photography; TPMKA, triple-phase multimodal knowledge aggregation; US, United States; W, diffuse white-light illumination.

**Abstract**

**Purpose:** To develop and evaluate an open-source multimodal artificial intelligence framework for bacterial-versus-fungal microbial keratitis classification and examine whether pooling geographically distinct populations creates shortcuts that inflate performance.

**Design:** Retrospective, multicenter, multicountry diagnostic accuracy study.

**Participants:** 1,645 patients with bacterial or fungal microbial keratitis from hospitals in the United States and India, contributing 17,158 slit-lamp photographs acquired with diffuse blue-light, sclerotic-scatter, and diffuse white-light illumination.

**Methods:** The framework combined cross-modality contrastive learning, modality-specific fine-tuning, and learned feature fusion with 7 clinical variables for patient-level prediction. Reference diagnoses used smear, culture, and polymerase chain reaction testing when available, with clinician adjudication of culture-negative cases after expert chart review. The framework and 5 comparators shared 5 patient-disjoint outer folds, stratified by country and organism class, and separate inner validation. Secondary analyses compared overall and within-country performance, included a country-only prevalence reference, and evaluated input ablations and imbalance-mitigation strategies.

**Main Outcome Measures:** Patient-level macro-F1, class-specific F1, accuracy, and area under the receiver operating characteristic curve (AUROC), overall and by country.

**Results:** The proposed framework achieved the highest overall macro-F1 (83.06%; 95% confidence interval [CI], 81.58–84.56) and accuracy (84.50%; 95% CI, 83.16–85.84) among matched comparators, with AUROC 89.9% (95% CI, 88.7–91.1). Paired comparisons numerically favored the framework but did not consistently reach statistical significance. Within-country AUROC was 83.9% in India and 67.0% in the United States, with bacterial F1 of 47.78% in India and fungal F1 of 40.97% in the US. The country-only reference achieved 81.88% overall accuracy and 78.2% AUROC, despite 50.0% AUROC within each country. Resampling and group distributionally robust optimization improved some locally underrepresented-group outcomes without eliminating the geographic pattern.

**Conclusions:** The proposed open-source framework integrates complementary slit-lamp images and clinical information and achieved favorable overall classification relative to matched comparators. Its evaluation also showed how geographic differences in organism frequencies can make multicountry performance more reassuring than within-country discrimination. These findings support further development of multimodal diagnostic assistance with geographically independent clinical validation.

Microbial keratitis (MK) is a potentially sight-threatening infection for which timely identification of the causative organism can guide treatment.[1,2] Yet clinical appearance can be ambiguous when differentiating organism types,[3-5] and laboratory confirmation is often delayed or unavailable.[4,5] Artificial intelligence that combines routinely acquired ocular images with clinical findings could help clinicians make earlier, better-informed treatment decisions while confirmatory testing is pending. Slit-lamp photography (SLP) offers complementary illumination modalities of the infection highlighting key morphology: blue light with fluorescein highlights epithelial defects, sclerotic scatter reveals intracorneal abnormalities, and white light documents broader ocular anatomy and pathology.[6,7] Integrating images with key clinical risk factors from the patient's history[3] offers an opportunity to extend diagnostic assistance across settings with different levels of specialist and laboratory access.

Deep learning approaches to keratitis classification include convolutional networks, vision transformers, and multimodal systems,[8–12] and the growing literature has prompted a systematic review and meta-analysis protocol.[13] Recent studies have explored fusion of image illumination modalities and incorporation of clinical information,[11,14] but integrating variable, sometimes incomplete image sets into a patient-level prediction remains challenging. We developed triple-phase multimodal knowledge aggregation (TPMKA), a novel, open-source framework that combines learning across a patient's images, adaptation to individual illumination modalities, and learned feature fusion with seven clinical variables. Building on our image-segmentation methods,[15,16] TPMKA accommodates variable image numbers and missing illumination modalities. Our primary objective was to develop and evaluate this framework against matched comparators in 1,645 patients with MK from India and the United States, providing a reproducible basis for diagnostic assistance that draws on the information collected during eye examinations.

Achieving this objective also requires understanding what a model has learned during training. Shortcut learning occurs when a model predicts a diagnosis partly from incidental features whose association with the diagnosis in the development data may not hold in a different clinical setting. For example, in MK, regional differences in bacterial and fungal prevalence may allow a model to exploit site-associated cues, including camera characteristics, illumination patterns, or other nonlesional image features, rather than infection morphology. Such associations can persist when different patients from the same sites appear in training and testing. Prior studies have identified site-associated label leakage and impaired cross-site generalization in keratitis models.[8,14] Our secondary objective was to clarify this risk using a country-only reference, within-country evaluation, input ablations, and imbalance-mitigation analyses. Together, these aims connect methodological progress with evaluation that can distinguish favorable multicountry results from diagnostic information likely to remain useful in individual clinical populations.

## Methods

### Study Population and Data Collection

We retrospectively analyzed clinical records and slit-lamp photographs from 1,645 patients with bacterial or fungal MK from four hospitals in the United States (US) and one in India. Patients were included through the prospective Automated Quantitative Ulcer Analysis (AQUA) study or retrospective cohorts from participating institutions. Inclusion and exclusion criteria for the main prospective AQUA study have been reported,[17] and the same eligibility criteria were applied to the retrospective cohorts. The cohort included 1,048 patients with fungal keratitis and 597 with bacterial keratitis, contributing 17,158 images. Each patient contributed one affected eye and baseline data were obtained at the initial presentation to a cornea specialist. We combined images and clinical variables from that encounter to produce one patient-level prediction. Throughout this report, country refers to India or the United States, and center refers to an individual participating hospital.

Reference diagnoses were based on smear, culture, and polymerase chain reaction (PCR) results when available. For cases without a documented positive laboratory result, the lead cornea specialist at each site assigned the reference diagnosis after expert chart review and consultation with the treating clinician. Clinicians were instructed to exclude cases that remained ambiguous because of multiple treatments or other confounding factors. This composite reference standard allowed the inclusion of some cases without positive laboratory findings. The analysis used the diagnostic labels in the dataset frozen on December 5, 2024. This label set remained fixed throughout model development and evaluation so that all models, data splits, and comparative analyses used the same reference labels. Of the 1,645 patients, 962 (58.5%) had a documented positive laboratory or culture result, 296 (18.0%) had no documented positive laboratory or culture result, and information on laboratory or culture positivity was unavailable for 387 (23.5%).

Images were acquired using up to three illumination modalities: diffuse blue light with topical fluorescein staining (B), sclerotic scatter (ScS), and diffuse white light (W). The number of images varied across patients and modalities, and some patients lacked images from one or more modalities, reflecting routine clinical acquisition. Figure 1B shows a representative patient set, and Table 1 and Figure 1A summarize the dataset. Clinical variables included country, contact-lens use, trauma history, organic-material exposure, water exposure, and presenting corrected and uncorrected visual acuity of the affected eye (Table 2).

The study was conducted under protocols reviewed by the University of Michigan Institutional Review Board (HUM00174923, prospective Automated Quantitative Ulcer Analysis [AQUA] study; HUM00198258, retrospective Automated Quantitative Ulcer Analysis [retroAQUA]), the Duke University Institutional Review Board (Pro00064195), and the Aravind Medical Research Foundation Institutional Ethics Committee. Data from outside institutions were obtained under data use agreements held by the University of Michigan, with the relevant institutional review board documentation maintained on file. Informed consent was obtained for the prospective AQUA and Aravind studies. The remaining studies were determined to be exempt by the relevant review boards, which waived the requirement for informed consent because the research posed minimal risk to participants; images and clinical metadata were de-identified before analysis. The study was conducted in accordance with the Declaration of Helsinki. All available eligible cases were included; no a priori sample-size calculation was performed.

**Model Development and Patient-Level Prediction**

TPMKA was designed to integrate complementary slit-lamp illumination modalities – diffuse blue light with fluorescein staining, sclerotic scatter, and diffuse white light – with clinical information through three learning phases. Phase One, cross-modality contrastive learning, encouraged agreement between representations of images from the same patient. Phase Two, modality-specific fine-tuning, adapted the image networks separately for each illumination modality to retain classification-relevant information. Phase Three used a learned multilayer perceptron (MLP)-based feature-fusion model to combine the fixed modality-specific representations and produce one prediction per patient. The modality-specific networks were frozen during this phase, and only the MLP-based feature-fusion model was optimized. The image encoder was the b1 variant of Pyramid Vision Transformer version 2 (PVT_v2_b1).[18] Figures 2 and 3 illustrate the training framework and network architecture. Technical specifications and training objectives are provided in Appendix 1 (available at https://www.aaojournal.org).

The seven clinical variables were recorded during image collection.[17] For the binary history variables, 1 denoted positive and −1 denoted negative. Missing values across all variables were coded as 0. All images were cropped using binary limbus masks generated automatically by our previously developed limbus-segmentation approach.[15,16] Thirty photographs from 23 patients were excluded because the predicted limbus masks contained no nonzero pixels; all affected patients retained other usable photographs. The masks delineated the limbus and contained no lesion-specific or organism-label information. Features were averaged across a patient's images within each illumination modality before fusion. Zero feature vectors represented missing illumination modalities, allowing patients with incomplete image sets to remain in the patient-level analysis.

**Comparator Models**

We compared TPMKA with DeepIK,[10] ResNet-50,[19] DenseNet-121,[20] SwinV2-T,[21] and PVT_v2_b1[18] implementations adapted to the present binary task. All methods shared the same preprocessing steps, clinical variables, and patient partitions. Each baseline used a separate network for each illumination modality and combined modality-specific predictions by patient-level voting; TPMKA used learned feature fusion. Probability aggregation and implementation details are provided in Appendix 1. These comparisons assess the implementations evaluated in this cohort and do not reproduce each published system's original clinical evaluation.

**Validation and Statistical Analysis**

We evaluated classification using five patient-disjoint outer folds stratified jointly by country and organism class. We kept all images from a patient in a single fold. Within each outer-training partition, 10% of patients were reserved for similarly stratified inner validation, yielding approximately 72% for training, 8% for validation, and 20% for testing. Methods shared the inner splits for checkpoint and model selection, and each outer test fold was evaluated after those choices were fixed. This protocol assessed new patients from the participating populations; external validation at an unseen center or in an unseen country was outside the scope of this study.

Phase One checkpoints were selected by the lowest inner-validation contrastive loss. Subsequent phases and supervised baselines used a country-and-class equity-scaled F1 criterion that rewarded overall performance while penalizing disparities between country-class groups. This group-aware checkpoint selection was shared across the Original and imbalance-mitigation settings. The exact criterion and training settings are described in Appendix 1.

Patient-level outcomes included accuracy, fungal and bacterial F1, macro-F1, and area under the receiver operating characteristic curve (AUROC), with bacterial keratitis as the positive class for AUROC and a fixed classification threshold of 0.5. Macro-F1 was the mean of the two class-specific F1 scores. Unless otherwise noted, we calculated each metric separately in each held-out fold and reported the arithmetic mean of the five values. Overall performance therefore refers to evaluating both countries together within each test fold, followed by averaging across folds.

We calculated 95% percentile confidence intervals (CI) from 10,000 patient-level bootstrap samples, resampling within fixed outer-fold, country, and organism-class strata and recomputing the mean of the five fold metrics. Paired intervals for between-model differences used the same sampled patients for both models and were unadjusted for multiple comparisons. Patient-level correctness was also compared using two-sided exact McNemar tests, with Holm correction for the five baseline comparisons and a nominal significance level of 0.05. Confidence intervals describe patient-sampling uncertainty conditional on the fitted models and observed country-class composition. Fungal and bacterial sensitivities were additionally reported overall and within each country. Additional statistical details are provided in Appendix 1.

**Secondary Country and Imbalance Analyses**

To examine the contribution of country to overall performance, we included a country-only prevalence reference that used no images or individual clinical findings. It assigned the same score to every patient within a country and classified all US patients as bacterial and all Indian patients as fungal, corresponding to the more common organism in each country. We compared its overall and within-country performance with PVT_v2_b1 voting and TPMKA. Because this reference cannot discriminate between patients within a country, it illustrates how country membership alone can support favorable aggregate metrics.

We evaluated the contribution of different inputs by retraining image-only and no-location variants of TPMKA under the shared patient-level protocol. Metadata-only logistic-regression models used either all seven variables or the six non-location variables. We also trained and tested the full framework separately within each country. These analyses assessed classification with restricted inputs or populations and whether the country-class pattern persisted.

To assess whether training adjustments improved performance for the locally less common organism, we compared the original setting with country-class resampling, group distributionally robust optimization (Group DRO),[22] and their combination. Resampling increased representation of underrepresented country-class groups during training; Group DRO reweighted classification losses across those groups. The original setting used empirical-risk minimization for fitting and shared group-aware checkpoint selection. We also evaluated a balanced subset of 348 patients, with 87 in each country-class group. All 87 US-fungal patients were included. For each of the other three country-class groups, 87 patients were randomly selected after restricting eligibility to patients with all three illumination modalities. Because balancing also reduced sample size and changed case composition, that analysis could not isolate the effect of prevalence balance.

## Results

The Indian cohort included 961 fungal cases (82.0%) and 211 bacterial cases (18.0%), whereas the US cohort included 87 fungal cases (18.4%) and 386 bacterial cases (81.6%) (Table 1).

### Overall Classification and Comparator Performance

TPMKA achieved overall macro-F1 of 83.06% (95% CI, 81.58–84.56), accuracy of 84.50% (95% CI, 83.16–85.84), and AUROC of 89.9% (95% CI, 88.7–91.1). Bacterial and fungal F1 were 78.14% and 87.99%, respectively. Table 3 reports the matched comparison with DeepIK, ResNet-50, DenseNet-121, SwinV2-T, and PVT_v2_b1. Held-out-fold receiver operating characteristic (ROC) curves are shown in Figure 4 (available at https://www.aaojournal.org).

The proposed MLP-based feature-fusion model outperformed all five baselines on overall fungal F1, bacterial F1, macro-F1, accuracy, and AUROC point estimates (Table 3). Relative to PVT_v2_b1 voting, the strongest baseline on overall macro-F1 and accuracy, the gains of our TPMKA model were 1.47 percentage points in bacterial F1, 0.84 in macro-F1, and 0.55 in accuracy.

The proposed models also had the highest point estimates for all five metrics within each illumination modality (Table 3). The numerical gains over the strongest modality-matched comparator were greatest for blue-light images: 2.87 percentage points in bacterial F1, 1.73 in macro-F1, and 1.10 in accuracy, each relative to PVT_v2_b1.

**Performance Within Countries and the Country-Only Reference**

Country-specific evaluation revealed lower discrimination than the overall estimate: TPMKA AUROC was 83.9% (95% CI, 81.0–86.7) in India and 67.0% (95% CI, 59.9–73.8) in the United States (Table 4; Figure 5). F1 was 47.78% for bacterial cases in India and 40.97% for fungal cases in the United States, the locally less common organism in each country. PVT_v2_b1 voting had higher US AUROC (70.5% versus 67.0%) and macro-F1 (67.38% versus 65.98%) than the proposed model. At the fixed 0.5 threshold, fungal sensitivity in the US was 30.13% (95% CI, 20.85–39.48), and bacterial sensitivity in India was 38.88% (95% CI, 32.24–45.54).

The country-only reference achieved overall macro-F1 of 79.36%, accuracy of 81.88%, and AUROC of 78.2%. Within each country, its AUROC was 50.0%, and F1 for India-bacterial and US-fungal cases was 0% (Table 4).

The complete framework achieved higher overall macro-F1 (83.06% vs 81.32%) and AUROC (89.9% vs 88.7%) than the image-only variant. India-bacterial and US-fungal F1 were 43.43% and 34.86% for the image-only variant and 46.03% and 36.10% after removing explicit country information from the complete framework. Thus, the geographic pattern persisted with both input restrictions. These comparisons describe the effects of the evaluated configurations but do not identify the specific image or patient characteristics used for prediction (Table 5).

Training the framework separately within each country did not resolve the difficulty classifying the locally less common organism. Metadata-only and single-country results are provided in Table 5.

***Effects of Imbalance Mitigation***

Resampling improved India-bacterial F1 from 47.78% to 52.92% and US-fungal F1 from 40.97% to 44.67%, with overall macro-F1 of 82.23% compared with 83.06% in the original setting. Group DRO yielded corresponding India-bacterial and US-fungal F1 scores of 50.67% and 44.29%, respectively, and overall macro-F1 of 82.43% (Table 6). Combining resampling with Group DRO yielded India-bacterial F1 of 50.59%, US-fungal F1 of 40.30%, and overall macro-F1 of 80.99%. This combination did not consistently improve on either strategy alone.

In the balanced subset of 348 patients, with 87 in each country-class group, TPMKA achieved overall macro-F1 of 70.23% and accuracy of 70.43%. This analysis changed sample size, patient composition, and organism frequencies, so it cannot isolate the effect of prevalence balance. Full model comparisons are provided in Table 7 (available at https://www.aaojournal.org). The evaluated strategies improved some outcomes for locally underrepresented organisms, with modest reductions in overall macro-F1, but did not eliminate the geographic pattern. These comparisons are descriptive and involve overlapping populations.

**Additional Analyses**

Configuration comparisons are reported in Tables 8 and 9 (available at https://www.aaojournal.org). Paired bacterial-F1 differences numerically favored TPMKA across all five comparisons. Unadjusted 95% intervals excluded zero against ResNet-50 and DenseNet-121; the difference versus PVT_v2_b1 was 1.47 percentage points (95% CI, −0.52 to 3.45). No McNemar accuracy comparison remained statistically significant after Holm adjustment (Table 10, available at https://www.aaojournal.org).

Patient-level bootstrap confidence intervals for the six primary models, overall and within each country, are provided in Table 11 (available at https://www.aaojournal.org). For TPMKA in the US, the 95% CIs were 59.75% to 71.38% for macro-F1 and 82.04% to 86.68% for accuracy. Patient and image distributions in the Original and Balanced datasets, including patient-level availability of each illumination modality by country and organism class, are summarized in Table 12 (available at https://www.aaojournal.org).

Within-country sensitivities and corresponding out-of-fold confusion counts, overall and by country, are reported in Table 13 (available at https://www.aaojournal.org). Resampling increased India-bacterial sensitivity from 38.88% to 50.75% and US-fungal sensitivity from 30.13% to 41.63%, while India-fungal sensitivity decreased from 94.59% to 91.78% and US-bacterial sensitivity from 96.63% to 90.93%.

**Exploratory Visualization**

Gradient-weighted class activation mapping (Grad-CAM) examples from held-out folds showed activation over lesions and the central cornea, as well as the sclera, eyelid margin, or background (Figure 6). These visualizations identify regions for further investigation but do not establish whether those regions or geography-associated cues determined the predictions.

**Discussion**

We developed TPMKA, a novel open-source multimodal framework for bacterial versus fungal MK classification that integrates complementary slit-lamp illumination modalities with clinical information. Its three-phase approach combines learning across a patient's images, adaptation to individual illumination modalities, and MLP-based feature fusion. In matched comparisons, TPMKA numerically outperformed the five comparator implementations on overall fungal F1, bacterial F1, macro-F1, accuracy, and AUROC, achieving an overall macro-F1 of 83.06%, accuracy of 84.50%, and AUROC of 89.9%. Corresponding estimates across the comparator implementations ranged from 86.62% to 87.76%, 75.32% to 76.67%, 81.01% to 82.22%, 82.67% to 83.95%, and 88.3% to 89.8%, respectively. These findings support the framework as a promising approach to etiologic classification and provide a basis for further evaluation in clinical populations.

The framework addresses practical features of slit-lamp photography: different illumination modalities reveal complementary findings, patients contribute varying numbers of images, and some illumination modalities may be unavailable. The training sequence brings these inputs together while retaining modality-specific learning before fusion. Earlier keratitis studies have also combined images with clinical information, as in Prajna's multimodal models,[14] or fused illumination modalities, as in Won's image-fusion approach[11] and KeraFusionNet's modality experts.[23] TPMKA builds on this work by integrating cross-modality representation learning, modality-specific fine-tuning, and patient-level feature aggregation within one framework. Its inclusion of incomplete image sets is practically relevant, although robustness to different acquisition and missing-data patterns requires further testing.

The input comparison also supports evaluating clinical information alongside images. The complete framework achieved higher overall macro-F1 and AUROC than the image-only variant (83.06% versus 81.32% and 89.9% versus 88.7%, respectively). Relative to PVT_v2_b1 voting, the strongest comparator on overall macro-F1 and accuracy, the gains were 0.84 percentage points in macro-F1 and 0.55 percentage points in accuracy. The 0.55-percentage-point difference in accuracy corresponds to approximately 5 to 6 additional correct classifications per 1,000 patients in this cohort. The paired bacterial-F1 interval included zero for this comparator, and no accuracy comparison was significant after Holm correction. These favorable overall estimates warrant further evaluation, with uncertainty about the magnitude of improvement over the strongest comparator. Differences in cohorts, reference standards, prediction units, and external testing also prevent ranking TPMKA against published systems by headline accuracy alone.

A second key contribution is the geographic analysis, which makes the clinical meaning of overall performance more explicit. Although TPMKA's overall AUROC was 89.9%, within-country AUROC was 83.9% in India and 67.0% in the US. The locally less common organisms were particularly challenging, with bacterial F1 of 47.78% in India and fungal F1 of 40.97% in the US. Country membership alone achieved 81.88% overall accuracy and an AUROC of 78.2%, despite providing no discrimination within either country. The country-only reference therefore directly illustrates a shortcut opportunity: knowing where a patient was recruited can predict the predominant organism without recognizing the infection in that patient's eye.

A learned classifier could benefit from the same association through explicit location or image features that convey country or site. In a mixed-country ROC analysis, country information can help distinguish a bacterial case from the US from a fungal case from India, even when it provides no discrimination within either country. If a model relies on this association, its predictions may become unreliable where organism frequencies differ. Patient-disjoint testing does not remove the opportunity when each fold retains the same geographic pattern. Center- and acquisition-associated cues have produced analogous problems in other medical imaging settings.[24,25] Thus, a favorable overall score can accurately describe a mixed cohort while providing an overly reassuring picture of performance in an individual clinic. The overall-to-country gap does not, by itself, measure how much TPMKA relied on a shortcut.

These analyses extend prior keratitis research on generalization by making the contribution of geographic composition explicit. Redd and colleagues recognized site-associated label leakage during multicenter development and subsequently used single-site training with separate-site validation.[8] Prajna and colleagues reported that a trial-trained image model achieved AUROC of 95% on trial test data but 70% on prospectively representative data, discussing possible acquisition cues.[14] External-site evaluation has also been central to keratitis screening[9] and DeepIK.[10] Esmaili and colleagues evaluated subtype classification in an independent hospital cohort,[26] while Jeong and Jeong's patient-disjoint study reported modest discrimination.[27] Our contribution is to connect a simple country-only reference with within-country, input-ablation, and mitigation analyses under a shared design. This makes the interpretive problem visible even when a model performs favorably against matched comparators, without attributing differences among published results to any single cause.

The input analyses identify practical directions for improving the framework and investigating shortcut reliance. The country-class pattern persisted both without explicit location and with images alone, so removing a location field does not establish independence from geography. Image appearance, clinical characteristics, and missingness may still convey site information. These results support testing that possibility; they do not prove that TPMKA inferred country from photographs. Disease spectrum, referral patterns, and sample size remain alternative or contributing explanations. Similarly, Grad-CAM activation outside the lesion highlights regions for investigation but does not establish dependence on them. Tests that alter acquisition cues while preserving disease content, together with evaluation across independent settings, are needed to distinguish these explanations.

For clinicians, local prevalence is useful because it informs the pretest probability of infection. The challenge is to determine whether a model adds information that distinguishes the organism in an individual patient. TPMKA achieved the highest overall macro-F1, accuracy, and AUROC point estimates among the matched comparators. Nevertheless, lower within-country than pooled AUROC was observed across all five comparator models (Table 11), and low F1 for locally less common organisms was also evident in the DeepIK and PVT_v2_b1 comparisons (Table 7). This pattern therefore extends beyond the proposed framework. In the original setting, TPMKA's overall accuracy of 84.50% coexisted with misclassification of approximately 70% of US fungal cases and 61% of India bacterial cases at the evaluated threshold (Table 13), precisely where diagnostic assistance may be especially valuable. F1 depends on prevalence and the decision threshold and should not be interpreted as sensitivity. Adjusting the threshold changes the tradeoff between fungal and bacterial sensitivity without changing AUROC; the US AUROC of 67.0% also indicates limited discrimination beyond the particular choice of a 0.5 cutoff. These findings highlight a broader challenge in multicountry diagnostic evaluation: favorable pooled scores, including numerical improvements over established approaches, do not establish reliable recognition of locally uncommon infections. Reporting performance within countries and organism classes makes these limitations visible. The observed pattern does not, by itself, establish that geographic shortcuts caused the errors.

The same risk can arise in studies of retinal photographs, optical coherence tomography, and other ophthalmic images. If patients with a target disease are recruited mainly from a referral center and controls mainly from a screening program, devices, acquisition protocols, or image-processing patterns may become associated with the diagnosis. Patient-disjoint testing that preserves these associations can reward recognition of the data source and make pooled performance appear more clinically informative than it is within either population. Large or multicenter datasets alone therefore do not establish generalizability. Reporting within-site and disease-group performance, alongside testing in independent clinical and acquisition settings, helps determine whether an apparent advance remains useful when the setting changes. These safeguards should accompany the development of ophthalmic classifiers, especially when disease prevalence differs substantially across contributing sites.

Resampling increased India-bacterial F1 from 47.78% to 52.92% and US-fungal F1 from 40.97% to 44.67%; Group DRO produced corresponding values of 50.67% and 44.29%. These numerical improvements were accompanied by lower overall macro-F1. Combining the strategies did not consistently improve on either approach alone. The balanced subset showed higher F1 for these groups, but it also changed prevalence, sample size, and case composition, preventing attribution of the differences to balancing alone. The additional sensitivity summaries showed that resampling produced the largest gains for the locally less common infections (India-bacterial, 38.88% to 50.75%; US-fungal, 30.13% to 41.63%) but reduced sensitivity for the locally more common infections, whereas Group DRO produced smaller rare-class sensitivity gains while better preserving US-bacterial sensitivity and the combined strategy was not consistently additive.

Several limitations should guide the next stage of evaluation. The cohort included patients classified at the December 5, 2024 label freeze as having bacterial or fungal keratitis and excluded viral, parasitic, polymicrobial, noninfectious, and indeterminate cases. Reference diagnoses combined laboratory testing with clinical adjudication, potentially introducing misclassification or incorporation bias if model inputs also informed adjudication. Diagnoses can also change with the disease course, treatment response, or subsequent testing. A later comparison with the updated study database identified four patients whose diagnoses were revised to polymicrobial or noninfectious keratitis; we retained them with their original labels so that all models, data splits, and comparisons used the same reference labels. Excluding them from the saved out-of-fold predictions, without retraining, changed overall accuracy and macro-F1 by less than 0.1 percentage points. The present results therefore concern a defined binary task and do not establish performance across the full differential diagnosis or the effect of model assistance on treatment decisions and outcomes.

Geographic evaluation was limited to one Indian center and four US centers, with only 87 US-fungal cases. These cohort proportions are not estimates of population prevalence, and country-level results may conceal differences among centers. Because the cohort had also informed model development, these internally validated estimates require confirmation in independently collected clinical populations. Missing modalities may reflect acquisition patterns and provide an opportunity for shortcut learning. In the Balanced subset, only US-fungal cases could lack an illumination modality; accordingly, this descriptive sensitivity analysis cannot isolate the effect of prevalence balance. Transportability and the contribution of geographic cues therefore remain uncertain.

The next step is to determine whether TPMKA improves diagnostic decisions in independent, prospectively collected clinical populations. Evaluation should include sites with different organism distributions and acquisition workflows, reporting overall and local discrimination alongside class sensitivity, predictive values, and calibration. A simple prevalence reference can help distinguish information associated with population membership from the additional information supplied by the model. Public release of the matching implementation and analysis code will enable independent reproduction and further development of the framework.

TPMKA provides an open-source, three-phase multimodal approach to bacterial versus fungal microbial keratitis classification and achieved the highest overall macro-F1, accuracy, and AUROC among matched comparators. More broadly, the comparison showed that favorable pooled performance across established approaches can conceal limited recognition of locally uncommon infections and that the evaluated mitigation strategies only partly alleviate this problem. Reporting country- and organism-specific performance and validating in geographically independent populations are therefore essential before clinical deployment.

**Data availability**

The study images are not publicly available because of institutional patient privacy restrictions and data-use agreements. Access may be requested from the corresponding author by email and is subject to appropriate approvals and data-use agreements with the University of Michigan.

**Code availability**

The open-source code is available at https://github.com/yqwang01/TPMKA. The implementation and analysis code corresponding to the reported experiments will be released in the same repository.

**Author contributions**

Yiqing Wang: Writing - review & editing, Writing - original draft, Visualization, Validation, Software, Methodology, Investigation, Formal analysis, Conceptualization. Maria A. Woodward: Writing - review & editing, Validation, Supervision, Resources, Project administration, Investigation, Funding acquisition, Data curation, Conceptualization. Ziyun Yang: Software, Formal analysis, Data curation. N. Venkatesh Prajna: Writing - review & editing, Supervision, Resources, Data curation. Chunming He: Software, Formal analysis, Writing – review & editing. Leslie M. Niziol: Writing – review & editing, Validation, Investigation, Formal analysis, Data curation. Mercy Pawar: Validation, Resources, Data curation. Ming-Chen Lu: Formal analysis, Data curation, Writing – review & editing. Guillermo Amescua: Resources, Data curation. Rachel Wozniak: Resources, Data curation. Sejal Amin: Resources, Data curation. Abinaya Krishnan: Resources, Data curation. Prabhleen Kochar: Resources, Data curation. Sina Farsiu: Writing – review & editing, Validation, Supervision, Resources, Project administration, Investigation, Funding acquisition, Data curation, Conceptualization. All authors reviewed and approved the final manuscript.

**Declaration of Generative AI and AI-assisted Technologies in the Writing Process**

During preparation of this work, the authors used OpenAI ChatGPT and Codex to assist with language editing, manuscript organization, formatting, and methodological review. The authors reviewed and revised the resulting suggestions and take full responsibility for the accuracy and integrity of the manuscript.

**Figure Legends**

Figure 1. Multicountry cohort and a representative patient with fungal keratitis. (A) Slit-lamp image counts by illumination modality and patient distributions by country and organism. Country bars are normalized to 100%; labels give counts and within-country percentages. (B) Blue-light, sclerotic-scatter, and white-light photographs from the same patient, with seven corresponding clinical variables. US indicates United States; logMAR, logarithm of the minimum angle of resolution.

Figure 2. Three-phase training framework for patient-level microbial keratitis classification. Phase One, cross-modality contrastive learning, learns representations by contrasting independently augmented images from the same patient against images from other patients. Phase Two, modality-specific fine-tuning, initializes and fine-tunes a separate network for each illumination modality using classification and contrastive objectives. Clinical variables enter the image-level networks in both phases. In Phase Three, an MLP-based feature-fusion model averages the resulting joint image-clinical features within each modality, concatenates the three modality vectors, and applies a fusion multilayer perceptron (MLP) trained with classification loss to produce one bacterial-versus-fungal prediction per patient. Missing modalities are represented by zero vectors. Photographs illustrate inputs and are reproduced from the representative case in Figure 1. B indicates blue-light illumination; ScS, sclerotic-scatter illumination; W, white-light illumination.

Figure 3. Image-clinical network architecture and contrastive-learning design. (A) An image encoder and a linear clinical-variable embedding feed a joint-feature multilayer perceptron (MLP). Separate heads support classification and contrastive learning. Phase One, cross-modality contrastive learning, uses the contrastive objective; Phase Two, modality-specific fine-

tuning, uses both objectives. Joint features preceding the heads supply the separate Phase Three fusion MLP in Figure 2. (B) Schematic contrastive objective. Independently augmented images from the same patient form a positive pair; sampled images from other patients provide negative examples regardless of diagnosis. Positive pairs need not come from different illumination modalities. Photographs are reproduced from the representative case in Figure 1 to illustrate inputs, not specific training augmentations. The diagram illustrates the training objective, not an observed embedding. PVT_v2_b1 indicates the b1 variant of Pyramid Vision Transformer version 2; L_cls, classification loss; L_con, contrastive loss; z, projected representation.
Figure 5. Overall and within-country discrimination for TPMKA, PVT_v2_b1 voting, and the country-only prevalence reference. Points show the arithmetic mean of five held-out-fold AUROCs. Error bars show patient-level bootstrap 95% confidence intervals for the learned models; no confidence interval is supplied for the country-only reference. Overall estimates evaluate both countries together within each test fold. The country-only reference achieves favorable overall discrimination but has an AUROC of 50.0% within each country because every patient in that country receives the same score. These comparisons demonstrate the influence of geographic composition on evaluation; they do not establish which features the learned models used. The dashed line marks an AUROC of 50.0%. TPMKA indicates triple-phase multimodal knowledge aggregation; PVT, Pyramid Vision Transformer; AUROC, area under the receiver operating characteristic curve; CI, confidence interval. PVT_v2_b1 denotes the b1 variant of Pyramid Vision Transformer version 2.
Figure 6. Gradient-weighted class activation mapping (Grad-CAM) examples from the modality-specific networks. Each illumination modality is shown as an original image and its Grad-CAM overlay. Maps use each patient's held-out-fold Phase Two checkpoint and target the class

predicted by the final Phase Three ensemble. Redder regions indicate larger Grad-CAM values for the selected class's modality-specific score. The maps describe localization for that score and do not directly explain the ensemble, establish country inference, or demonstrate causal shortcut use. Rows include correctly and incorrectly classified cases with all three illumination modalities. Predict / Truth gives the ensemble prediction and reference label; 0 indicates fungal keratitis and 1 indicates bacterial keratitis.

## Tables

Multimodal Knowledge Aggregation for Microbial Keratitis: Geographic Confounding and Shortcut Risk in Pooled Multicountry Data

**Table 1. Distribution of patients and slit-lamp photographs by country and organism class.**

| Country | Patient Count | | | Blue Light Image | | | Sclerotic Scatter Image | | | White Light Image | | |
|---|---|---|---|---|---|---|---|---|---|---|---|---|
| | Fungal | Bacterial | All | Fungal | Bacterial | All | Fungal | Bacterial | All | Fungal | Bacterial | All |
| India | 961 | 211 | 1,172 | 3,546 | 778 | 4,324 | 3,644 | 847 | 4,491 | 3,653 | 805 | 4,458 |
| United States | 87 | 386 | 473 | 208 | 930 | 1,138 | 206 | 925 | 1,131 | 288 | 1,328 | 1,616 |
| Overall | 1,048 | 597 | 1,645 | 3,754 | 1,708 | 5,462 | 3,850 | 1,772 | 5,622 | 3,941 | 2,133 | 6,074 |

The cohort comprised 1,645 patients and 17,158 images.

**Table 2. Clinical variables included in the multimodal framework.**

| Attribute Name | Description |
|---|---|
| Location | The country where the patient received the diagnosis. Possible values: India, US. |
| Trauma | Indicates whether the patient has experienced trauma to the eye. Possible values: Yes, No, Unknown. |
| Organic Matter | Indicates whether the patient's eye has been exposed to organic matter. Possible values: Yes, No, Unknown. |
| Water Exposure | Indicates whether the patient's eye has been exposed to water. Possible values: Yes, No, Unknown. |
| Contact Lens Usage | Indicates whether the patient wears contact lenses. Possible values: Yes, No, Unknown. |
| Presenting Corrected Visual Acuity | The patient's visual acuity in the affected eye, in logMAR units, with the correction worn at presentation, if any. |
| Presenting Uncorrected Visual Acuity | The patient's visual acuity in the affected eye, in logMAR units, without any correction. |

logMAR indicates logarithm of the minimum angle of resolution; US, United States. Encoding and preprocessing details are described in Appendix 1.

**Table 3. Full performance matrix for modality-specific and patient-level models.**

| Method | Modality/Aggregation | F1 Score | | | Accuracy | AUROC |
|---|---|---|---|---|---|---|
| | | Fungal | Bacterial | Macro-F1 | | |
| ResNet-50 | B | 86.24 (1.41) | 66.38 (4.58) | 76.31 (2.72) | 80.52 (1.93) | 84.06 (2.69) |
| | ScS | 85.35 (1.67) | 71.08 (2.75) | 78.21 (2.18) | 80.56 (2.07) | 86.80 (2.47) |
| | W | 84.72 (1.98) | 72.71 (3.09) | 78.72 (2.39) | 80.44 (2.30) | 85.83 (2.05) |
| | Voting | 86.86 (1.31) | 75.32 (2.93) | 81.09 (2.10) | 82.86 (1.81) | 88.49 (1.80) |
| DenseNet-121 | B | 84.80 (2.27) | 65.22 (5.40) | 75.01 (3.65) | 78.88 (3.03) | 81.38 (3.48) |
| | ScS | 86.95 (1.03) | 73.43 (2.49) | 80.19 (1.56) | 82.52 (1.31) | 88.07 (2.15) |
| | W | 84.35 (0.35) | 72.09 (1.27) | 78.22 (0.79) | 79.95 (0.59) | 85.47 (2.15) |
| | Voting | 86.62 (0.68) | 75.39 (2.37) | 81.01 (1.52) | 82.67 (1.16) | 88.30 (2.14) |
| SwinV2-T | B | 85.80 (1.70) | 65.77 (1.33) | 75.78 (0.98) | 79.98 (1.69) | 84.60 (2.30) |
| | ScS | 87.67 (2.12) | 73.75 (3.08) | 80.71 (2.55) | 83.24 (2.55) | 88.86 (1.91) |
| | W | 85.95 (2.34) | 74.55 (3.44) | 80.25 (2.83) | 81.91 (2.74) | 87.15 (1.50) |
| | Voting | 87.38 (1.30) | 76.25 (2.08) | 81.82 (1.65) | 83.53 (1.57) | 89.14 (1.54) |
| DeepIK | B | 84.58 (2.91) | 66.71 (3.94) | 75.65 (3.35) | 78.96 (3.45) | 83.72 (2.87) |
| | ScS | 87.05 (0.94) | 72.92 (2.60) | 79.99 (1.53) | 82.51 (1.16) | 87.09 (1.44) |
| | W | 83.62 (1.32) | 71.24 (2.99) | 77.43 (1.98) | 79.15 (1.68) | 85.57 (2.15) |
| | Voting | 87.22 (1.94) | 76.51 (3.55) | 81.87 (2.64) | 83.47 (2.41) | 89.05 (2.34) |
| PVT_v2_b1 | B | 86.87 (1.89) | 67.62 (4.55) | 77.25 (3.08) | 81.34 (2.57) | 84.74 (2.41) |
| | ScS | 87.24 (1.95) | 73.43 (4.11) | 80.34 (2.94) | 82.78 (2.56) | 88.67 (2.21) |
| | W | 85.82 (1.52) | 73.68 (2.89) | 79.75 (1.99) | 81.60 (1.81) | 87.56 (2.22) |
| | Voting | 87.76 (1.31) | 76.67 (2.31) | 82.22 (1.76) | 83.95 (1.63) | 89.77 (1.57) |
| TPMKA | B | 87.47 (2.64) | 70.49 (4.04) | 78.98 (3.24) | 82.44 (3.25) | 85.92 (3.29) |
| | ScS | 87.97 (1.64) | 75.01 (2.63) | 81.49 (2.12) | 83.76 (2.04) | 89.52 (2.33) |
| | W | 86.12 (1.05) | 75.66 (1.85) | 80.89 (1.30) | 82.34 (1.18) | 88.52 (1.69) |

| Method | Modality/Aggregation | F1 Score | | | Accuracy | AUROC |
|---|---|---|---|---|---|---|
| | | Fungal | Bacterial | Macro-F1 | | |
| | Feature ensemble | 87.99 (1.87) | 78.14 (3.01) | 83.06 (2.43) | 84.50 (2.30) | 89.93 (1.99) |

Values are five-fold mean percentages with sample standard deviations in percentage points in parentheses. B indicates blue-light illumination; ScS, sclerotic-scatter illumination; W, white-light illumination; AUROC, area under the receiver operating characteristic curve. Modality-specific comparisons are descriptive; paired inferential analyses concern the final patient-level outputs (Table 10). Bootstrap confidence intervals are in Table 11.

**Table 4. Performance across and within countries for the country-only reference, PVT voting, and TPMKA.**

| Population | Method | Macro-F1 (%) | Accuracy (%) | AUROC (%) |
|---|---|---|---|---|
| Overall | Country only | 79.36 | 81.88 | 78.18 |
| Overall | PVT voting | 82.22 | 83.95 | 89.77 |
| Overall | TPMKA | 83.06 | 84.50 | 89.93 |
| India | Country only | 45.05 | 82.00† | 50.00 |
| India | PVT voting | 65.77 | 83.87 | 82.11 |
| India | TPMKA | 69.35 | 84.56 | 83.92 |
| US | Country only | 44.94 | 81.61† | 50.00 |
| US | PVT voting | 67.38 | 84.15 | 70.54 |
| US | TPMKA | 65.98 | 84.36 | 66.96 |

Unless marked †, values are means of five held-out-fold percentages. † Country-only within-country accuracies are derived from the complete-cohort majority-class counts: India, 961/1,172 = 82.00%; US, 386/473 = 81.61%. These two entries are descriptive cohort proportions, not regenerated fold-mean estimates. Overall results combine countries within each fold. F1 and accuracy use the fixed 0.5 cutoff; AUROC uses continuous scores. Patient-level bootstrap 95% confidence intervals for PVT and TPMKA are in Table 11 and Figure 5. The country-only score is constant within each country and fold, giving within-country AUROC of 50.00%. AUROC indicates area under the receiver operating characteristic curve; US, United States. PVT indicates Pyramid Vision Transformer; PVT voting uses the PVT_v2_b1 variant.

**Table 5. Input comparisons and country-restricted training.**

| Country | Input / training setting | F1 Score | | | Accuracy | AUROC |
|---|---|---|---|---|---|---|
| | | Fungal | Bacterial | Macro-F1 | | |
| Overall | TPMKA (image + metadata) | 87.99 (1.87) | 78.14 (3.01) | 83.06 (2.43) | 84.50 (2.30) | 89.93 (1.99) |
| India | | 90.93 (2.01) | 47.78 (9.20) | 69.35 (5.54) | 84.56 (3.30) | 83.92 (4.04) |
| US | | 40.97 (11.91) | 90.98 (1.52) | 65.98 (6.67) | 84.36 (2.68) | 66.96 (11.13) |
| Overall | TPMKA (image only) | 86.68 (1.68) | 75.97 (3.18) | 81.32 (2.43) | 82.86 (2.20) | 88.68 (1.95) |
| India | | 89.95 (1.73) | 43.43 (11.58) | 66.69 (6.63) | 82.94 (3.02) | 81.56 (5.12) |
| US | | 34.86 (12.58) | 89.98 (0.76) | 62.42 (6.60) | 82.67 (1.52) | 65.89 (11.48) |
| Overall | TPMKA (no country) | 87.59 (1.47) | 77.28 (2.53) | 82.44 (1.99) | 83.95 (1.85) | 89.78 (2.35) |
| India | | 90.81 (1.43) | 46.03 (8.98) | 68.42 (5.16) | 84.30 (2.44) | 84.04 (4.38) |
| US | | 36.10 (8.86) | 90.24 (1.14) | 63.17 (4.79) | 83.09 (1.91) | 63.85 (12.14) |
| Overall | Metadata only model | 86.60 (0.30) | 72.32 (0.66) | 79.46 (0.48) | 81.95 (0.41) | 81.76 (1.96) |
| India | | 90.14 (0.29) | 1.86 (2.55) | 46.00 (1.38) | 82.08 (0.51) | 59.01 (5.97) |
| US | | 0.00 (0.00) | 89.87 (0.31) | 44.94 (0.16) | 81.61 (0.51) | 62.48 (2.67) |
| Overall | Metadata only model without country | 84.46 (0.52) | 64.67 (1.33) | 74.57 (0.92) | 78.42 (0.74) | 78.84 (1.69) |
| India | | 89.94 (0.23) | 6.04 (3.77) | 47.99 (1.85) | 81.83 (0.37) | 57.55 (3.74) |
| US | | 15.46 (6.71) | 81.74 (1.92) | 48.60 (4.07) | 69.98 (2.90) | 52.48 (7.50) |
| India | TPMKA (India-only training) | 91.24 (1.00) | 45.08 (11.27) | 68.16 (6.13) | 84.90 (1.92) | 83.27 (3.56) |
| US | TPMKA (US-only training) | 16.48 (12.42) | 84.79 (7.37) | 50.63 (6.09) | 75.08 (10.05) | 58.04 (11.33) |

Values are five-fold mean percentages with sample standard deviations in percentage points in parentheses. F1 and accuracy use the fixed 0.5 cutoff; AUROC uses continuous scores. Country-restricted models were trained and tested only within the named country; these are not cross-country transfer evaluations. AUROC indicates area under the receiver operating characteristic curve; US, United States.

**Table 6. TPMKA performance across all five imbalance-analysis settings.**

| Setting | Population | Fungal F1 | Bacterial F1 | Macro-F1 | Accuracy | AUROC |
|---|---|---|---|---|---|---|
| ERM | Overall | 87.99 (1.87) | 78.14 (3.01) | 83.06 (2.43) | 84.50 (2.30) | 89.93 (1.99) |
| ERM | India | 90.93 (2.01) | 47.78 (9.20) | 69.35 (5.54) | 84.56 (3.30) | 83.92 (4.04) |
| ERM | US | 40.97 (11.91) | 90.98 (1.52) | 65.98 (6.67) | 84.36 (2.68) | 66.96 (11.13) |
| Resampling | Overall | 87.22 (0.79) | 77.24 (2.10) | 82.23 (1.34) | 83.65 (1.08) | 89.42 (1.98) |
| Resampling | India | 90.59 (0.87) | 52.92 (8.73) | 71.76 (4.43) | 84.39 (1.34) | 84.35 (4.09) |
| Resampling | US | 44.67 (11.94) | 89.11 (1.45) | 66.89 (6.66) | 81.83 (2.68) | 68.21 (10.44) |
| Group DRO | Overall | 87.15 (1.61) | 77.71 (3.27) | 82.43 (2.39) | 83.71 (2.13) | 89.41 (2.04) |
| Group DRO | India | 90.15 (1.40) | 50.67 (10.37) | 70.41 (5.74) | 83.62 (2.45) | 83.30 (4.21) |
| Group DRO | US | 44.29 (5.53) | 90.61 (1.54) | 67.45 (3.47) | 83.94 (2.46) | 65.62 (11.01) |
| Resampling + Group DRO | Overall | 86.21 (0.68) | 75.78 (1.86) | 80.99 (1.22) | 82.43 (0.99) | 89.48 (1.34) |
| Resampling + Group DRO | India | 89.84 (1.02) | 50.59 (9.28) | 70.22 (4.94) | 83.19 (1.72) | 84.35 (3.09) |
| Resampling + Group DRO | US | 40.30 (13.06) | 88.36 (2.17) | 64.33 (7.32) | 80.56 (3.53) | 69.16 (9.81) |
| Balanced | Overall | 69.27 (6.37) | 71.18 (5.99) | 70.23 (5.60) | 70.43 (5.48) | 75.82 (5.89) |
| Balanced | India | 77.02 (4.71) | 75.44 (9.19) | 76.23 (6.79) | 76.44 (6.51) | 81.86 (6.84) |
| Balanced | US | 59.96 (15.80) | 67.44 (8.53) | 63.70 (11.88) | 64.35 (11.23) | 68.06 (8.71) |

Values are five-fold mean percentages with sample standard deviations in percentage points in parentheses. ERM indicates empirical-risk minimization; Group DRO, group distributionally robust optimization; AUROC, area under the receiver operating characteristic curve; US, United States. ERM is the Original setting in the main text. All settings use the same group-aware inner-validation checkpoint-selection rule. F1 and accuracy use the fixed 0.5 cutoff. The Balanced subset includes 348 patients with 87 per country-class group and changes sample size and case composition as well as prevalence. Table 7 contains the full three-model comparison.

## Figures

Multimodal Knowledge Aggregation for Microbial Keratitis: Geographic Confounding and Shortcut Risk in Pooled Multicountry Data

Figure 1. Multicountry cohort and a representative patient with fungal keratitis. (A) Slit-lamp image counts by illumination modality and patient distributions by country and organism. Country bars are normalized to 100%; labels give counts and within-country percentages. (B) Blue-light, sclerotic-scatter, and white-light photographs from the same patient, with seven corresponding clinical variables. US indicates United States; logMAR, logarithm of the minimum angle of resolution.

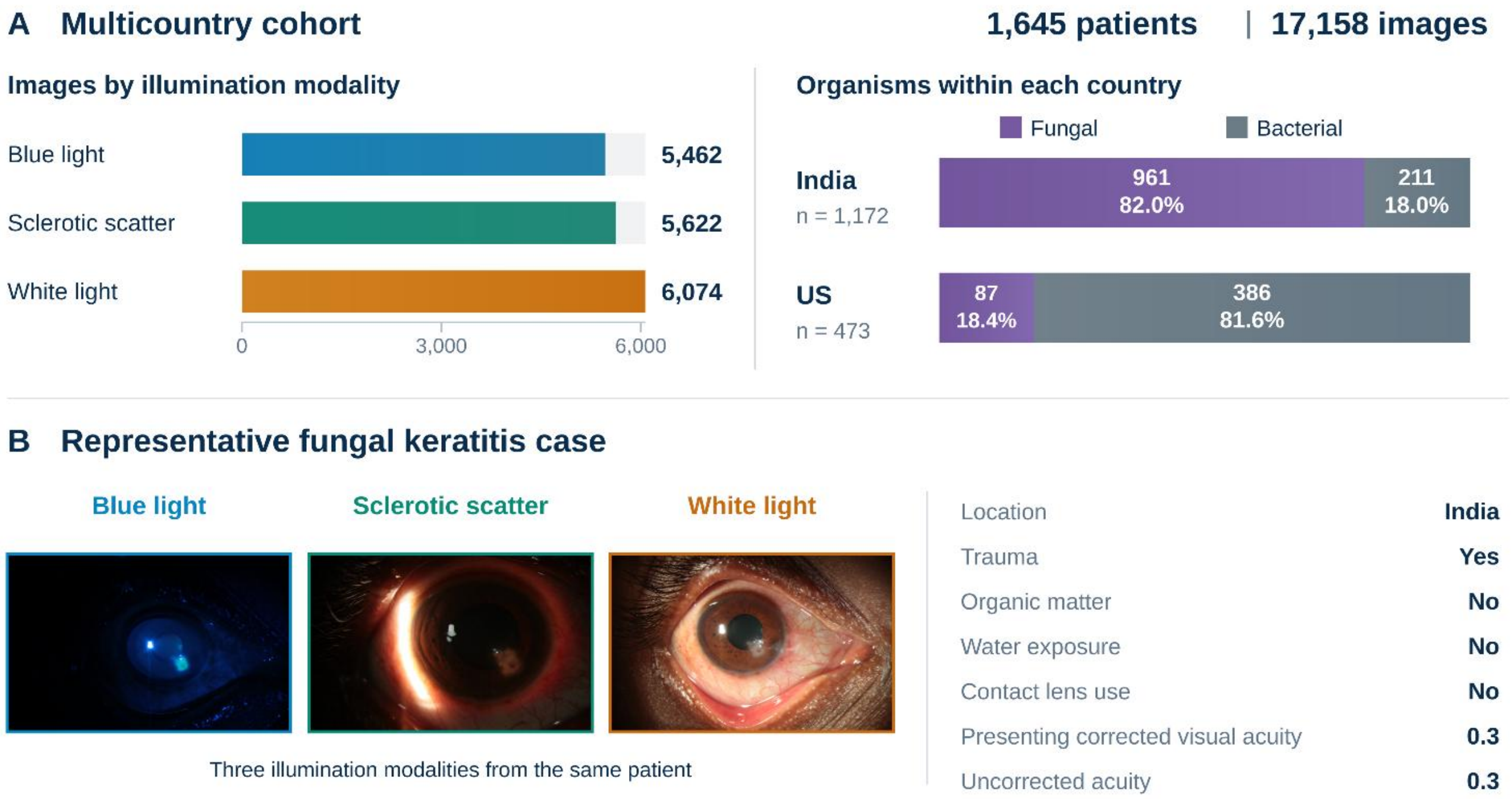

Figure 2. Three-phase training framework for patient-level microbial keratitis classification. Phase One, cross-modality contrastive learning, learns representations by contrasting independently augmented images from the same patient against images from other patients. Phase Two, modality-specific fine-tuning, initializes and fine-tunes a separate network for each illumination modality using classification and contrastive objectives. Clinical variables enter the image-level networks in both phases. In Phase Three, an MLP-based feature-fusion model averages the resulting joint image-clinical features within each modality, concatenates the three modality vectors, and applies a fusion multilayer perceptron (MLP) trained with classification loss to produce one bacterial-versus-fungal prediction per patient. Missing modalities are represented by zero vectors. Photographs illustrate inputs and are reproduced from the representative case in Figure 1. B indicates blue-light illumination; ScS, sclerotic-scatter illumination; W, white-light illumination.

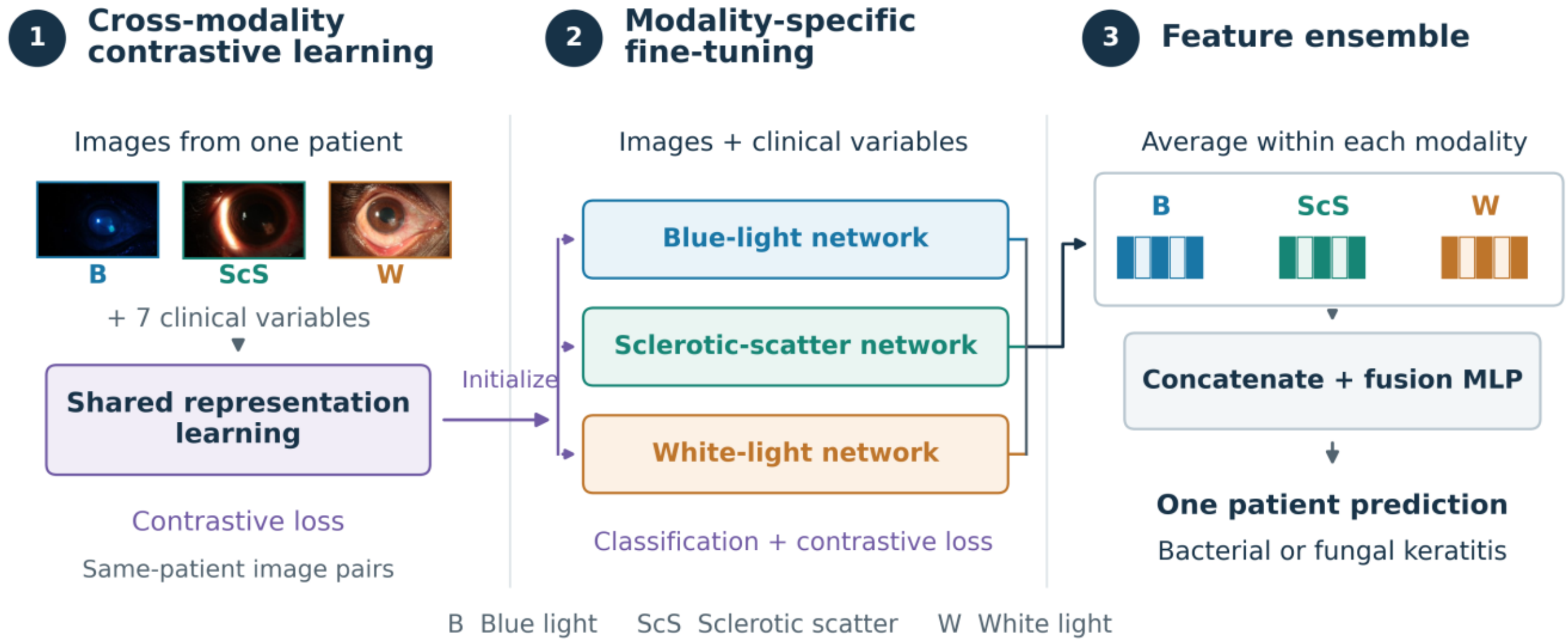

Figure 3. Image-clinical network architecture and contrastive-learning design. (A) An image encoder and a linear clinical-variable embedding feed a joint-feature multilayer perceptron (MLP). Separate heads support classification and contrastive learning. Phase One, cross-modality contrastive learning, uses the contrastive objective; Phase Two, modality-specific fine-tuning, uses both objectives. Joint features preceding the heads supply the separate Phase Three fusion MLP in Figure 2. (B) Schematic contrastive objective. Independently augmented images from the same patient form a positive pair; sampled images from other patients provide negative examples regardless of diagnosis. Positive pairs need not come from different illumination modalities. Photographs are reproduced from the representative case in Figure 1 to illustrate inputs, not specific training augmentations. The diagram illustrates the training objective, not an observed embedding. PVT_v2_b1 indicates the b1 variant of Pyramid Vision Transformer version 2; L_cls, classification loss; L_con, contrastive loss; z, projected representation.

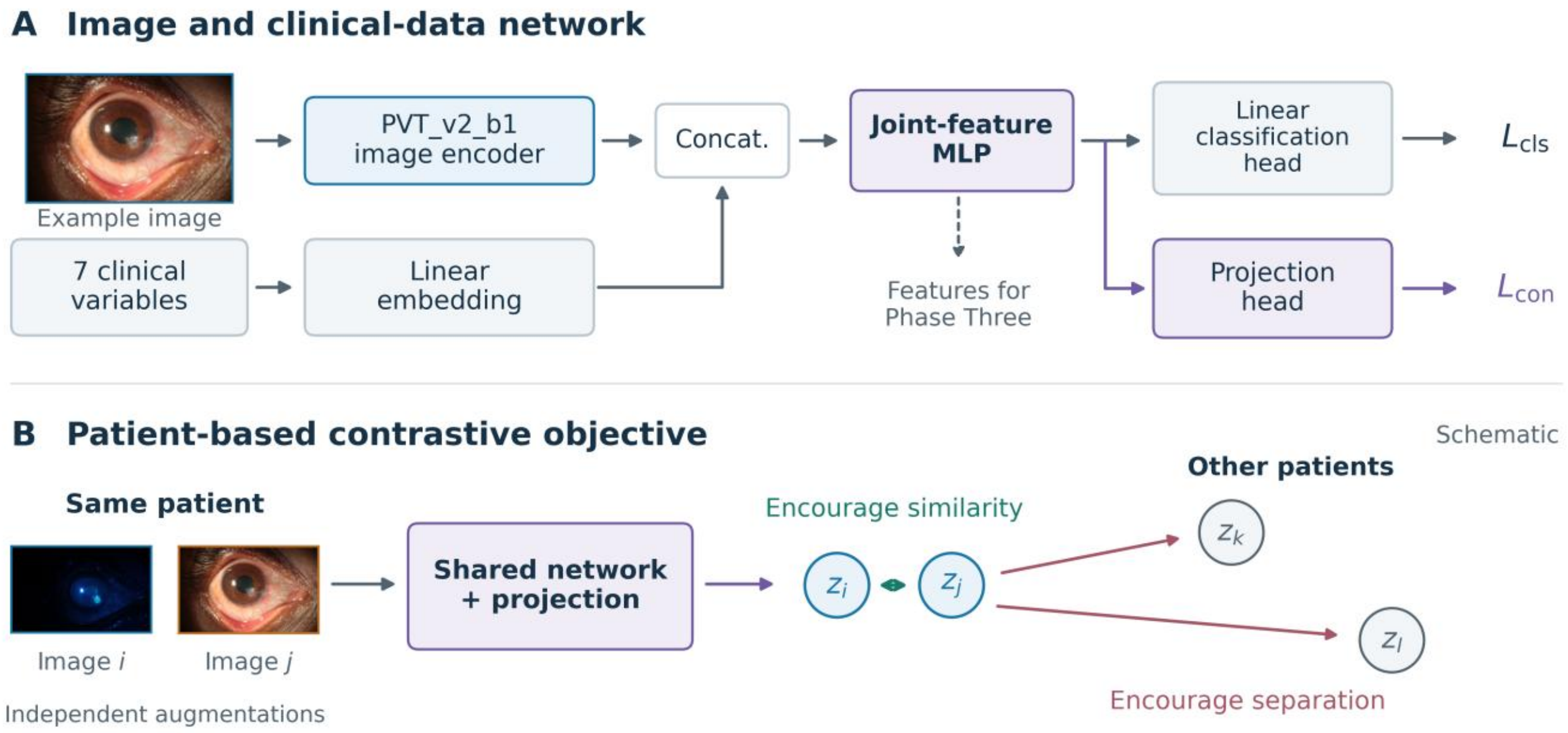

Figure 5. Overall and within-country discrimination for TPMKA, PVT_v2_b1 voting, and the country-only prevalence reference.

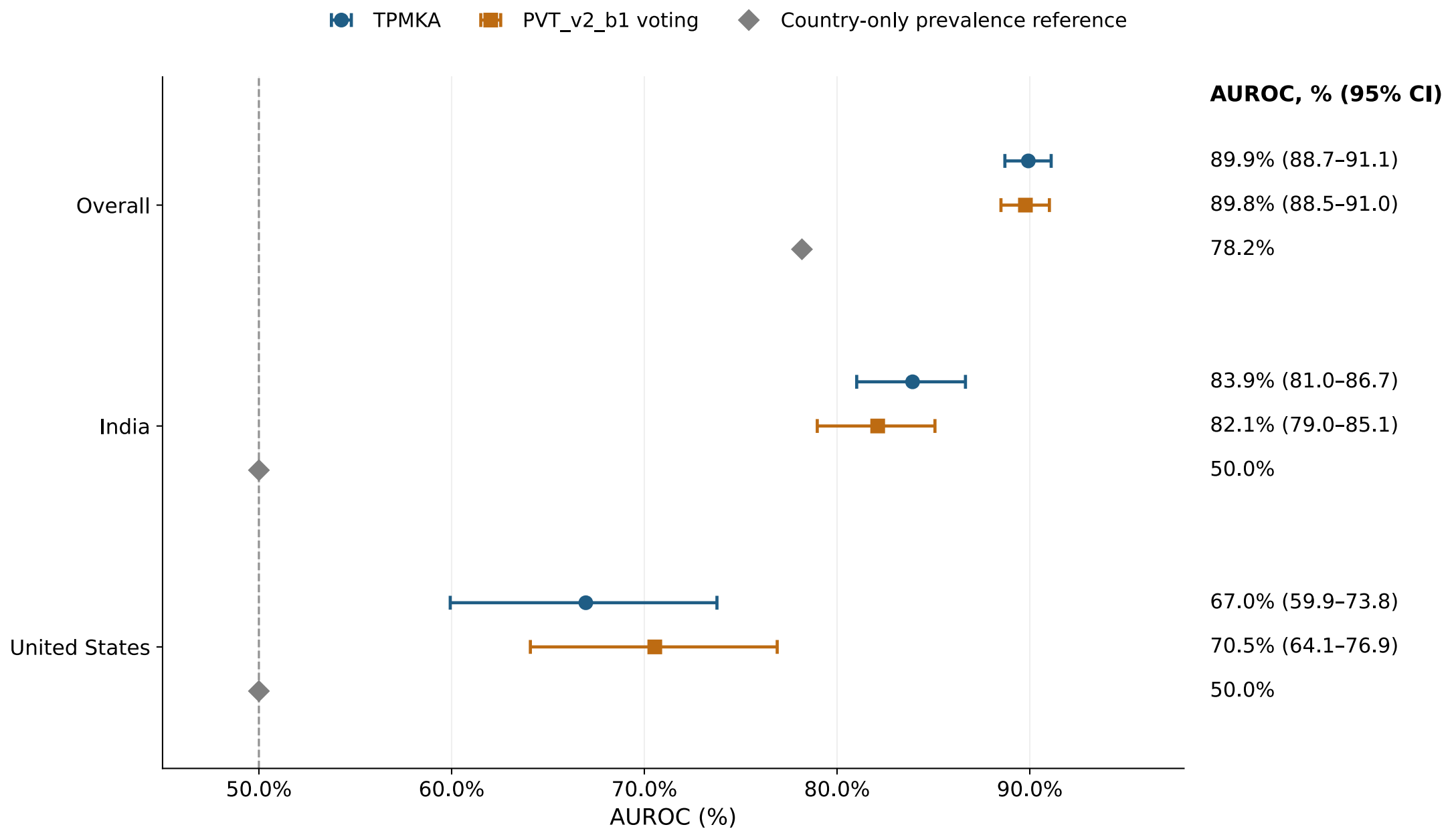

Points show the arithmetic mean of five held-out-fold AUROCs. Error bars show patient-level bootstrap 95% confidence intervals for the learned models; no confidence interval is supplied for the country-only reference. Overall estimates evaluate both countries together within each test fold. The country-only reference achieves favorable overall discrimination but has an AUROC of 50.0% within each country because every patient in that country receives the same score. These comparisons demonstrate the influence of geographic composition on evaluation; they do not establish which features the learned models used. The dashed line marks an AUROC of 50.0%. TPMKA indicates triple-phase multimodal knowledge aggregation; PVT, Pyramid Vision Transformer; AUROC, area under the receiver operating characteristic curve; CI, confidence interval. PVT_v2_b1 denotes the b1 variant of Pyramid Vision Transformer version 2.

Figure 6. Gradient-weighted class activation mapping (Grad-CAM) examples from the modality-specific networks. Each illumination modality is shown as an original image and its Grad-CAM overlay. Maps use each patient's held-out-fold Phase Two checkpoint and target the class predicted by the final Phase Three ensemble. Redder regions indicate larger Grad-CAM values for the selected class's modality-specific score. The maps describe localization for that score and do not directly explain the ensemble, establish country inference, or demonstrate causal shortcut use. Rows include correctly and incorrectly classified cases with all three illumination modalities. Predict / Truth gives the ensemble prediction and reference label; 0 indicates fungal keratitis and 1 indicates bacterial keratitis.

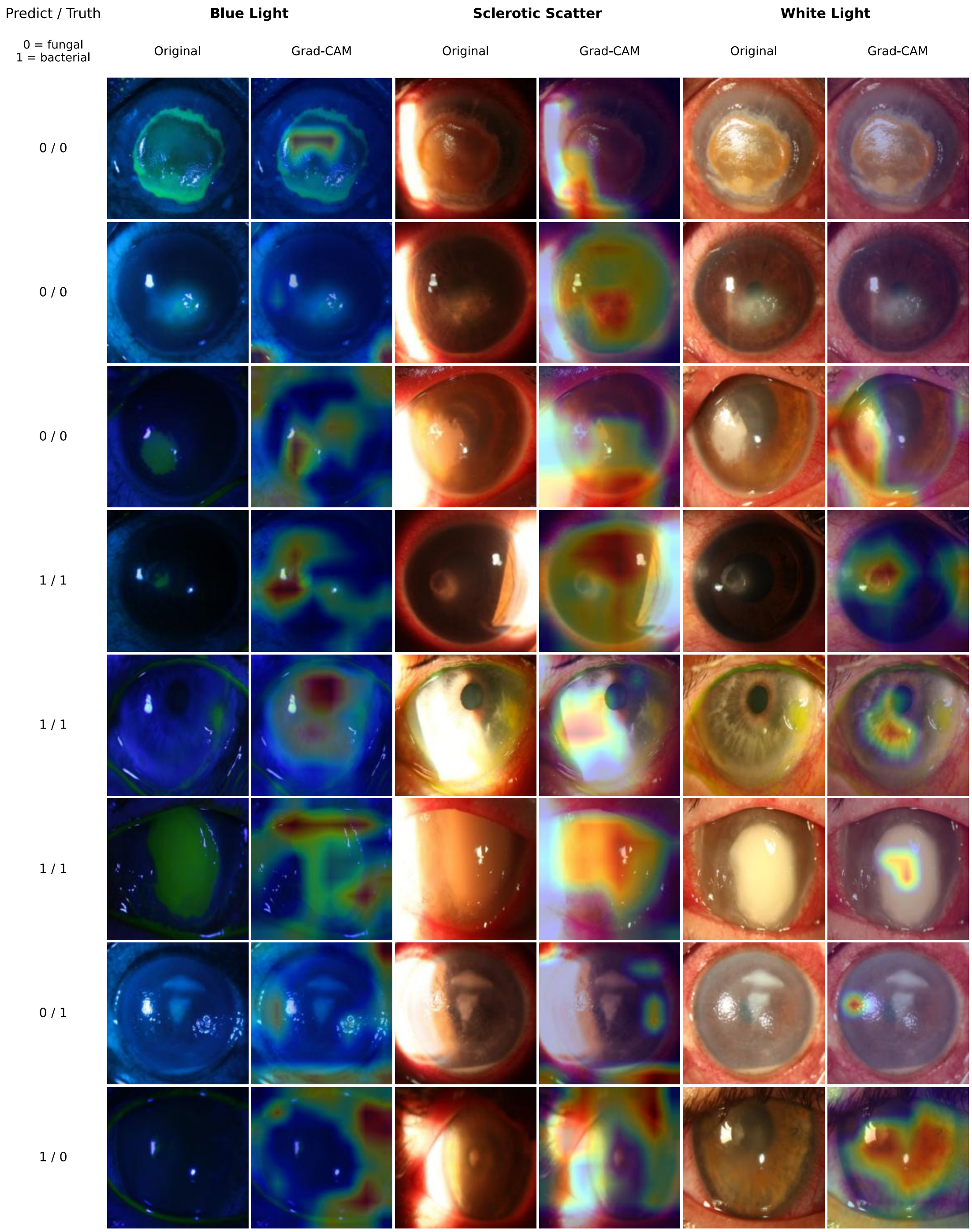

## Online supplementary materials

Multimodal Knowledge Aggregation for Microbial Keratitis: Geographic Confounding and Shortcut Risk in Pooled Multicountry Data

### Appendix 1 Supplementary methods

The following methods describe the model implementation, patient-level evaluation, and configuration comparisons supporting the clinical analyses in the main manuscript. Unless otherwise noted, overall estimates combine the countries within each held-out fold and are summarized as means of the five fold-specific metrics.

### Image preprocessing

We cropped images around the limbus, the boundary between the cornea and sclera, to reduce extracorneal background. Following our previous work,[1] we used the Bilateral Connectivity Network (BiconNet)[2] to localize the limbus from interpixel connectivity and generate a bounding box for each image. These bounding boxes defined the corneal crops used for model training. Binary limbus masks were generated automatically for all images using the previously developed segmentation model. The masks delineated the limbus/corneal boundary and contained no lesion-specific or organism-label information. Images for which automatic segmentation did not produce a usable mask were excluded before the final analytic cohort was constructed. A limbus mask was considered unusable when it contained no nonzero pixels, indicating that no valid limbal region had been detected. Mask usability was assessed programmatically by counting nonzero pixels in each predicted mask. Thirty photographs from 23 patients met this criterion and were excluded from the cropped-image dataset. All affected patients retained usable photographs, and no patient was excluded solely because of an unusable limbus mask. The contribution of cropping was evaluated in the ablation analysis (Table 9). Training augmentation included random cropping, flipping, rotation, and color jittering.

### Image encoders and clinical variables

We evaluated four image-encoder architectures: the convolutional neural networks ResNet-50[3] and DenseNet-121,[4] and the hierarchical transformers SwinV2-T[5,6] and Pyramid Vision Transformer (PVT).[7,8] These architectures provided modality-specific classifiers and patient-level voting baselines for comparison with TPMKA. PVT_v2_b1[7] was the backbone of TPMKA; its voting baseline had the highest overall macro-F1, accuracy, and area under the receiver operating characteristic curve (AUROC) among the five baseline methods (Table 3).

Following our previous study,[9] the framework incorporated seven clinical variables recorded during image collection: presenting corrected and uncorrected visual acuity of the affected eye, trauma history, organic-material exposure, country, water exposure, and contact-lens use. The contribution of these variables was evaluated in the metadata ablation analysis (Table 9). For trauma, organic-material exposure, water exposure, and contact-lens use, 1 denoted positive, −1 denoted negative, and 0 denoted missing; missingness was retained as an explicit category. Missing values occurred in 246, 781, 344, and 43 patients, respectively. Country was complete in the analytic cohort. Missing values for corrected and uncorrected visual acuity were imputed as 0 logMAR (20/20). All preprocessing parameters were fitted within the corresponding training fold and then applied to the validation and test data.

The PVT_v2_b1 encoder[7] extracted image features, and a linear layer embedded the clinical metadata. The two feature vectors were concatenated and passed through a multilayer perceptron (MLP) to form a joint representation. A projection head mapped this representation to the latent space used for contrastive learning, and a linear classification head produced the class scores.

**Phase One cross-modality contrastive learning**

Contrastive learning[10] encourages a representation to be more similar to its positive partner than to negative examples. A common approach constructs a positive pair from two independently augmented views of the same image and uses views of other images in the batch as negatives.[10,11]

We extended this approach to pairs of images from the same patient, aiming to learn representations shared across acquisition conditions and modalities.[12] At each iteration, we sampled $N$ patients and randomly selected two images per patient to form a positive pair. Sampled images from other patients served as negative examples.

Let $x_i$ and $x_j$ denote two images from the same patient. After independent random augmentation, the encoder $f(\cdot)$ mapped them to feature vectors $h_i$ and $h_j$. A projection head then produced latent vectors $z_i$ and $z_j$. The contrastive loss for this pair was

$$L_{con}=-log\frac{e^{\mathrm{sim}(z_i,z_j)/\tau}}{\sum_{k=1}^{2N}1_{k\neq i}e^{\mathrm{sim}(z_i,z_k)/\tau}} \quad (S1)$$

where $\mathrm{sim}(\cdot,\cdot)$ denotes cosine similarity and $\tau$ is the temperature parameter. The objective encouraged agreement between a patient's sampled images relative to images from other patients. Because patient identity also carries acquisition-related and patient-specific information, this objective does not by itself ensure disease-specific features or invariance to country. Table 8 reports the cross-modality contrastive learning (CMCL) configuration comparisons.

**Phase Two modality-specific fine-tuning**

We initialized three modality-specific networks with the Phase One pretrained parameters and fine-tuned each network using images from one illumination modality. This allowed each network to adapt the shared representation to the visual information available under its corresponding illumination condition.

Fine-tuning combined a cross-entropy classification loss (Equation S2) with the contrastive loss from Equation S1. In Equation S2, $y_i$ is the one-hot reference label for class i, $p_i$ is the predicted probability of that class, and C is the number of classes. The combined objective in Equation S3 used $\lambda = 0.1$, assigning weights of 0.9 to cross-entropy and 0.1 to the contrastive term.

$$L_{cls}=-\sum_{i=1}^{C}y_i log(p_i) \quad (S2)$$

$$L_{finetune}=(1-\lambda)L_{cls}+\lambda L_{con} \quad (S3)$$

**Phase Three MLP-based feature-fusion model**

For the baselines, predictions from the available modalities were combined by hard majority voting. Within each modality, bacterial softmax probabilities were averaged across all available images and thresholded at 0.5 to obtain a modality-level label. Ties in the patient-level vote were resolved by comparing the mean probability across available modalities with 0.5. The median modality-level probability served as the continuous score for receiver operating characteristic (ROC) analysis. Missing modalities were excluded. This approach combines modality-level predictions, whereas feature-level fusion can retain information that is lost when features are reduced to class labels.[13,14]

The proposed MLP-based feature-fusion model averaged features from the layer preceding each modality-specific classification head across that patient's images. The three resulting D-dimensional vectors were concatenated and passed to an MLP trained with cross-entropy loss (Equation S2). The concatenated 1,536-dimensional feature vector was processed by an MLP with dimensions 1,536→256→32→2, with ReLU activation after each hidden layer and dropout set to 0. A missing modality was represented by a zero vector. This produced one prediction per patient while allowing the number of images and the available modalities to vary. During Phase Three, the three modality-specific networks were frozen. Fixed 512-dimensional feature vectors were extracted without gradient updates, and only the MLP-based feature-fusion model was optimized.

### Cross-validation and checkpoint selection

All experiments used a shared five-fold outer cross-validation design with patient-disjoint folds stratified jointly by country and organism class. All images from a patient remained in the same fold. Within each outer-training partition, 10% of patients were reserved for an inner validation set using the same stratification, yielding approximately 72% of patients for training, 8% for inner validation, and 20% for outer testing. The same inner split was used across methods for training monitoring, checkpoint selection, hyperparameter selection, and model selection. The outer test fold was evaluated after these choices had been fixed.

Phase One checkpoints were selected by the lowest deterministic contrastive loss on the inner validation set. Phase Two, Phase Three, and supervised baseline checkpoints were selected by the highest country-and-class equity-scaled F1 on that set (Equation S4). This criterion divided inner macro-F1 by one plus the sum of the absolute differences between inner macro-F1 and each country-specific class F1, with $\alpha = 1$. $M$ is overall inner-validation macro-F1 and $F_{c,k}$ is the class-specific F1 for country c and class k. All F1 values were expressed on the 0–1 scale, and $\alpha$ was fixed at 1.0 for Phase Two, Phase Three, supervised baselines, ERM, and all retained imbalance-mitigation settings. Phase One used deterministic contrastive loss instead. For experiments using the complete 1,645-patient cohort, the inner-validation set in each outer fold comprised 132 patients: 77 India-fungal, 17 India-bacterial, 7 US-fungal, and 31 US-bacterial. Thus, the criterion considered both overall validation performance and variation across country-class groups.

$$ES - F1 = \frac{M}{1 + \alpha \sum_{c,k} \left| F_{c,k} - M \right|} \quad (S4)$$

### Training implementation and computational requirements

ResNet-50,[3] DenseNet-121,[4] SwinV2-T,[5] and PVT_v2_b1[7] were initialized with ImageNet-1k weights.[15] ResNet-50 and DenseNet-121 used ImageNet-pretrained weights from torchvision, whereas PVT_v2_b1 and SwinV2-T used their corresponding ImageNet-pretrained checkpoints from timm. The DeepIK comparator used a DenseNet-121 binary classifier trained separately for each illumination modality on 256 × 256 crops with weighted cross-entropy and AdamW; available modality-level predictions were combined by patient-level voting. Input images were cropped using automatically generated limbus masks as described above and resized to 256 × 256 pixels.

Except for the DeepIK comparator described above, models were trained with Adam[16] and a weight decay of $1 \times 10^{-3}$. Learning rates were $3 \times 10^{-5}$ in Phases One and Three and $1 \times 10^{-5}$ in Phase Two. Training lasted 150, 100, and 100 epochs, respectively, with a batch size of 32 and a learning-rate schedule comprising linear warm-up followed by polynomial decay. All models were implemented in PyTorch 2.5.1[17] and trained on a single NVIDIA GeForce RTX 4090 graphics processing unit (GPU).

Training a modality-specific PVT_v2_b1 model with a batch size of 32 required approximately 4.72 gibibytes (GiB) of peak CUDA-allocated memory and 5.07 GiB of peak CUDA-reserved memory. Median GPU computation time at inference was 13.17 ms per patient. Mean latency for the timed model pipeline was 50.33 ms per patient with a warm file cache and 56.22 ms after best-effort cache eviction. The timed pipeline began with loading previously generated limbus-cropped NPZ images and included NPZ reading, PIL conversion, resizing and normalization, host-to-GPU transfer, image encoding, within-modality feature aggregation, and ensemble prediction. It did not include automatic limbus segmentation, bounding-box generation, cropping of the original photographs, or conversion of the cropped images to NPZ files; these steps were performed beforehand.

### Performance measures and uncertainty

Performance was assessed using accuracy (Equation S5), fungal and bacterial F1 scores (Equation S6), macro-F1, and AUROC. Macro-F1 was the mean of the two class-specific F1 scores, and bacterial keratitis was the positive class for AUROC. In the equations, TP, TN, FP, and FN denote true positives, true negatives, false positives, and false negatives for the class being evaluated. The classification cutoff was fixed at 0.5. Unless otherwise noted, each metric was calculated separately in each held-out outer fold and summarized as the arithmetic mean of the five fold values. Sample standard deviations described variation between folds, and patient-level bootstrap confidence intervals quantified sampling uncertainty. Apart from the two derived country-only accuracies identified in Table 4, we did not pool predictions across folds to calculate these performance summaries.

$$\text{Accuracy} = \frac{TP+TN}{TP+TN+FP+FN} \quad \text{(S5)}$$

$$\text{F1-score} = 2\frac{Precision * Recall}{Precision+Recall} = \frac{2TP}{2TP+FP+FN} \quad \text{(S6)}$$

Uncertainty estimates were based on one out-of-fold prediction per patient. For each model, we generated 10,000 bootstrap samples by resampling patients with replacement within each fixed outer-fold, country, and organism-class stratum. In each replicate, we recomputed the metric separately in each fold and averaged the five values. We used the resulting distributions to calculate percentile 95% confidence intervals for AUROC, macro-F1, accuracy, fungal sensitivity, and bacterial sensitivity overall and within each country.

Between-model differences were estimated by resampling the same patients for both models and subtracting the comparator's metric from that of TPMKA. These paired bootstrap intervals were unadjusted for multiple comparisons. We also compared patient-level correctness across all 1,645 out-of-fold predictions using two-sided exact McNemar tests, with Holm correction for the five comparisons against baseline methods.

The nominal significance level was 0.05. Bootstrap intervals quantified patient-sampling variation conditional on the fitted cross-validation models and the observed country-class composition. They did not include uncertainty from repeating model training or drawing a new cohort with different country-class proportions. The five cross-validation training sets overlapped, so their fitted models were not independent. Fungal and bacterial sensitivities and pooled out-of-fold confusion counts are reported in Table 13.

### Input comparisons and geographic restriction

We compared the complete image-plus-metadata framework with image-only, no-country, metadata-only, and country-restricted configurations. A country-only prevalence reference quantified the discrimination available from country alone. Its score was constant among patients from the same country within a fold; consequently, it could rank patients from different countries but could not discriminate bacterial from fungal cases within either country. The country-only within-country accuracies marked † in Table 4 were derived from the cohort counts for majority-class prediction: 961/1,172 (82.00%) in India and 386/473 (81.61%) in the United States. These are complete-cohort descriptive proportions, rather than means of fold-specific accuracies; the other Table 4 entries retain the reported fold-mean estimator.

Two variants were retrained using the same patient-level five-fold protocol: an image-only model without metadata and a no-country model retaining the other six clinical variables. We also trained metadata-only logistic-regression models with either all seven variables or the six non-country variables. Within each training fold, the two logarithm of the minimum angle of resolution (logMAR) variables were median-imputed and standardized, and the categorical variables were one-hot encoded, with 0 retained as an explicit missing category for the binary clinical variables. Preprocessing parameters were estimated from training data only. Both metadata-only logistic-regression models used fixed, prespecified settings without hyperparameter selection: L2 regularization, C = 1.0, the liblinear solver, no class weighting, a maximum of 2,000 iterations, random_state = 0, and an intercept.

To evaluate geographic restriction, we trained and tested the complete three-phase pipeline separately in the India and United States (US) cohorts, which comprised 1,172 and 473 patients, respectively. Each model was evaluated only in its corresponding country. Table 5 reports the input ablations, metadata-only models, and country-specific training results.

### Country-class imbalance mitigation

We compared DeepIK, PVT_v2_b1 with patient-level modality voting, and the proposed MLP-based feature-fusion model under five settings: empirical-risk minimization (ERM), Resampling, group distributionally robust optimization (Group DRO), Resampling + Group DRO, and Balanced (Table 7). Table 6 reports all five settings for TPMKA. ERM retained the observed cohort distribution and used empirical-risk minimization for fitting, with the shared group-aware checkpoint-selection rule described above. Resampling used inverse-frequency weights for the four country-class groups and sampled with replacement, giving equal expected sampling frequencies to India-fungal, India-bacterial, US-fungal, and US-bacterial cases. Resampling was applied at all applicable training stages.

Group DRO[18] retained the original sampling distribution and reweighted classification losses for the four country-class groups. Cross-entropy losses were calculated separately for each country-class group within each minibatch. Adversarial group weights were initialized uniformly, updated by exponentiated gradient ascent with a step size of 0.01, and normalized to sum to one. Updates began after a five-epoch warm-up. Group DRO replaced the ordinary cross-entropy term during modality-specific fine-tuning and feature-ensemble training; the contrastive objectives were unchanged. The combined setting applied both weighted resampling and Group DRO.

Balanced used a subset of 348 patients, with 87 in each country-class group. All 87 US-fungal patients were retained. In each of the other three groups, 87 patients were randomly selected after restricting eligibility to patients with all three illumination modalities. The subset used newly assigned patient-disjoint five-fold splits rather than the Original-cohort fold assignments; all methods within the Balanced analysis used the same Balanced manifest and folds. Results were calculated at the fixed 0.5 cutoff and summarized as five-fold means and sample standard deviations overall and within each country. This subset tested performance under equal country-class representation. Because it also reduced cohort size and changed case composition, it could not isolate the effect of balancing from those of sample size and subset selection.

### Model visualization

We used gradient-weighted class activation mapping (Grad-CAM)[19] to visualize class-associated regions in the PVT_v2_b1[7] modality-specific networks. Maps were derived from the depthwise-convolution feature map in the MLP of the final block of the last PVT stage. Each image was processed with the Phase Two checkpoint from the patient's held-out outer fold. Across modalities, the target was the class predicted by the final Phase Three ensemble. Redder regions indicate larger Grad-CAM values for that class's Phase Two score. The maps therefore describe the modality-specific networks rather than the Phase Three fusion decision and do not establish causal feature importance.

### Configuration comparisons

Table 8 reports SwinV2-T and PVT_v2_b1 configurations labelled without or with cross-modality contrastive learning (CMCL), separately for each illumination modality and the patient-level output. Table 9 reports four configurations differing in metadata, limbus cropping, and contrastive regularization during fine-tuning. These comparisons are descriptive. A difference can be attributed to one component only when pretraining, initialization, aggregation, and all other training choices are held constant.

For SwinV2-T, the CMCL-labelled configuration increased blue-light macro-F1 from 75.78% to 78.72%, while sclerotic-scatter macro-F1 was essentially unchanged (80.71% versus 80.70%). Patient-level macro-F1 was 81.82% without CMCL and 82.20% with CMCL. For PVT_v2_b1, the corresponding modality-specific

macro-F1 values were 77.25% versus 78.98% for blue light, 80.34% versus 81.49% for sclerotic scatter, and 79.75% versus 80.89% for white light. Patient-level macro-F1 was 82.22% versus 83.06% (Table 8). The patient-level comparison does not isolate CMCL if the aggregation method also differs.

Among the four configurations in Table 9, the complete configuration achieved the highest macro-F1 and accuracy point estimates, at 83.06% and 84.50%, respectively. The configuration retaining metadata but omitting both limbus cropping and contrastive regularization achieved a macro-F1 of 80.96%. The image-only configuration retaining limbus cropping and the contrastive regularization achieved a macro-F1 of 81.32%. The configuration retaining metadata and limbus cropping but omitting the contrastive regularization achieved a macro-F1 of 82.52%. Because all four configurations used the same MLP-based feature-fusion model, rows 1 and 3 compared the effect of limbus cropping, rows 2 and 4 compared the effect of metadata, and rows 3 and 4 compared the effect of the paired contrastive regularization.

## Appendix references

**Supplementary tables**

**Table 7. Full comparison of country-class imbalance-mitigation settings.**

| Setting | Country | Method | F1 Score | | | Accuracy | AUROC |
|---|---|---|---|---|---|---|---|
| | | | Fungal | Bacterial | Macro-F1 | | |
| ERM | Overall | DeepIK | 87.22 (1.94) | 76.51 (3.55) | 81.87 (2.64) | 83.47 (2.41) | 89.05 (2.34) |
| | | PVT_v2_b1 | 87.76 (1.31) | 76.67 (2.31) | 82.22 (1.76) | 83.95 (1.63) | 89.77 (1.57) |
| | | TPMKA | 87.99 (1.87) | 78.14 (3.01) | 83.06 (2.43) | 84.50 (2.30) | 89.93 (1.99) |
| | India | DeepIK | 90.53 (1.96) | 46.93 (10.76) | 68.73 (6.13) | 83.96 (3.18) | 81.60 (5.51) |
| | | PVT_v2_b1 | 90.65 (1.07) | 40.88 (7.64) | 65.77 (4.22) | 83.87 (1.80) | 82.11 (3.42) |
| | | TPMKA | 90.93 (2.01) | 47.78 (9.20) | 69.35 (5.54) | 84.56 (3.30) | 83.92 (4.04) |
| | US | DeepIK | 40.36 (4.48) | 89.53 (1.21) | 64.94 (1.83) | 82.24 (1.58) | 68.72 (9.10) |
| | | PVT_v2_b1 | 44.01 (8.80) | 90.76 (0.87) | 67.38 (4.77) | 84.15 (1.59) | 70.54 (8.88) |
| | | TPMKA | 40.97 (11.91) | 90.98 (1.52) | 65.98 (6.67) | 84.36 (2.68) | 66.96 (11.13) |
| Resampling | Overall | DeepIK | 86.43 (1.16) | 76.91 (1.59) | 81.67 (1.28) | 82.92 (1.26) | 88.97 (2.09) |
| | | PVT_v2_b1 | 86.64 (1.14) | 76.72 (1.48) | 81.68 (1.19) | 83.04 (1.20) | 90.07 (1.28) |
| | | TPMKA | 87.22 (0.79) | 77.24 (2.10) | 82.23 (1.34) | 83.65 (1.08) | 89.42 (1.98) |
| | India | DeepIK | 89.58 (1.18) | 50.88 (3.69) | 70.23 (1.85) | 82.85 (1.56) | 83.11 (3.89) |
| | | PVT_v2_b1 | 90.11 (0.86) | 52.08 (4.27) | 71.09 (2.32) | 83.62 (1.27) | 84.77 (1.87) |
| | | TPMKA | 90.59 (0.87) | 52.92 (8.73) | 71.76 (4.43) | 84.39 (1.34) | 84.35 (4.09) |
| | US | DeepIK | 42.64 (12.18) | 90.07 (2.12) | 66.35 (6.98) | 83.10 (3.51) | 70.51 (15.38) |
| | | PVT_v2_b1 | 41.03 (11.59) | 89.06 (2.26) | 65.05 (6.50) | 81.61 (3.50) | 71.66 (6.63) |
| | | TPMKA | 44.67 (11.94) | 89.11 (1.45) | 66.89 (6.66) | 81.83 (2.68) | 68.21 (10.44) |

**Table 7 (continued). Full comparison of country-class imbalance-mitigation settings.**

| Setting | Country | Method | F1 Score | | | Accuracy | AUROC |
|---|---|---|---|---|---|---|---|
| | | | Fungal | Bacterial | Macro-F1 | | |
| Group DRO | Overall | DeepIK | 85.22 (0.93) | 74.16 (0.95) | 79.69 (0.63) | 81.22 (0.76) | 87.42 (1.43) |
| | | PVT_v2_b1 | 86.86 (1.51) | 76.46 (3.71) | 81.66 (2.53) | 83.16 (2.11) | 89.80 (1.81) |
| | | TPMKA | 87.15 (1.61) | 77.71 (3.27) | 82.43 (2.39) | 83.71 (2.13) | 89.41 (2.04) |
| | India | DeepIK | 88.78 (1.65) | 46.20 (4.84) | 67.49 (2.66) | 81.49 (2.28) | 79.96 (3.30) |
| | | PVT_v2_b1 | 90.78 (1.39) | 53.65 (10.60) | 72.22 (5.94) | 84.64 (2.49) | 83.44 (3.76) |
| | | TPMKA | 90.15 (1.40) | 50.67 (10.37) | 70.41 (5.74) | 83.62 (2.45) | 83.30 (4.21) |
| | US | DeepIK | 40.20 (9.37) | 88.37 (1.99) | 64.28 (5.48) | 80.56 (3.18) | 68.82 (10.42) |
| | | PVT_v2_b1 | 36.83 (12.02) | 87.70 (1.07) | 62.26 (6.13) | 79.49 (1.74) | 73.62 (8.78) |
| | | TPMKA | 44.29 (5.53) | 90.61 (1.54) | 67.45 (3.47) | 83.94 (2.46) | 65.62 (11.01) |
| Resampling + Group DRO | Overall | DeepIK | 85.46 (0.65) | 75.28 (1.31) | 80.37 (0.83) | 81.70 (0.73) | 87.95 (1.84) |
| | | PVT_v2_b1 | 86.12 (1.20) | 75.09 (1.73) | 80.60 (1.32) | 82.19 (1.32) | 89.83 (1.83) |
| | | TPMKA | 86.21 (0.68) | 75.78 (1.86) | 80.99 (1.22) | 82.43 (0.99) | 89.48 (1.34) |
| | India | DeepIK | 88.81 (0.90) | 48.43 (6.44) | 68.62 (3.22) | 81.66 (1.25) | 80.71 (2.80) |
| | | PVT_v2_b1 | 90.01 (1.04) | 49.82 (2.94) | 69.92 (1.70) | 83.36 (1.53) | 84.06 (3.02) |
| | | TPMKA | 89.84 (1.02) | 50.59 (9.28) | 70.22 (4.94) | 83.19 (1.72) | 84.35 (3.09) |
| | US | DeepIK | 41.21 (10.83) | 89.24 (0.91) | 65.22 (5.86) | 81.83 (1.80) | 69.36 (14.19) |
| | | PVT_v2_b1 | 37.64 (6.44) | 87.52 (1.96) | 62.58 (3.46) | 79.27 (2.74) | 70.60 (5.57) |
| | | TPMKA | 40.30 (13.06) | 88.36 (2.17) | 64.33 (7.32) | 80.56 (3.53) | 69.16 (9.81) |

**Table 7 (continued). Full comparison of country-class imbalance-mitigation settings.**

| Setting | Country | Method | F1 Score | | | Accuracy | AUROC |
|---|---|---|---|---|---|---|---|
| | | | Fungal | Bacterial | Macro-F1 | | |
| Balanced | Overall | DeepIK | 63.22 (10.05) | 71.59 (4.08) | 67.40 (6.78) | 68.14 (6.08) | 72.98 (7.13) |
| | | PVT_v2_b1 | 65.46 (2.80) | 70.64 (5.41) | 68.05 (3.80) | 68.39 (4.08) | 74.17 (4.10) |
| | | TPMKA | 69.27 (6.37) | 71.18 (5.99) | 70.23 (5.60) | 70.43 (5.48) | 75.82 (5.89) |
| | India | DeepIK | 70.80 (8.55) | 73.43 (3.33) | 72.11 (5.51) | 72.44 (5.05) | 76.80 (8.16) |
| | | PVT_v2_b1 | 70.36 (8.60) | 74.02 (7.71) | 72.19 (8.00) | 72.37 (7.95) | 79.25 (3.87) |
| | | TPMKA | 77.02 (4.71) | 75.44 (9.19) | 76.23 (6.79) | 76.44 (6.51) | 81.86 (6.84) |
| | US | DeepIK | 53.85 (11.79) | 69.91 (5.23) | 61.88 (8.01) | 63.83 (7.25) | 68.23 (8.28) |
| | | PVT_v2_b1 | 59.79 (8.46) | 67.16 (7.95) | 63.48 (6.48) | 64.42 (6.62) | 68.54 (9.32) |
| | | TPMKA | 59.96 (15.80) | 67.44 (8.53) | 63.70 (11.88) | 64.35 (11.23) | 68.06 (8.71) |

Values are five-fold mean percentages with sample standard deviations in percentage points in parentheses. F1 and accuracy use the fixed 0.5 cutoff; AUROC uses continuous scores. ERM is the Original setting in the main text. All retained settings use the same group-aware inner-validation checkpoint-selection rule. The Balanced subset changes both sample size and country-class composition and does not isolate the effect of balancing. ERM indicates empirical-risk minimization; Group DRO, group distributionally robust optimization; AUROC, area under the receiver operating characteristic curve; US, United States.

**Table 8. Configurations without or with cross-modality contrastive learning.**

| Backbone | Modality | CMCL | F1 Score | | | Accuracy |
|---|---|---|---|---|---|---|
| | | | Fungal | Bacterial | Macro-F1 | |
| SwinV2-T | B | | 85.80 (1.70) | 65.77 (1.33) | 75.78 (0.98) | 79.98 (1.69) |
| | | √ | 87.43 (1.47) | 70.02 (4.23) | 78.72 (2.78) | 82.30 (2.14) |
| | ScS | | 87.67 (2.12) | 73.75 (3.08) | 80.71 (2.55) | 83.24 (2.55) |
| | | √ | 87.76 (1.91) | 73.64 (2.61) | 80.70 (2.15) | 83.31 (2.23) |
| | W | | 85.95 (2.34) | 74.55 (3.44) | 80.25 (2.83) | 81.91 (2.74) |
| | | √ | 85.96 (1.36) | 75.45 (2.94) | 80.70 (2.09) | 82.15 (1.83) |
| | Ensemble | | 87.38 (1.30) | 76.25 (2.08) | 81.82 (1.65) | 83.53 (1.57) |
| | | √ | 87.69 (1.64) | 76.71 (3.47) | 82.20 (2.54) | 83.89 (2.22) |
| PVT_v2_b1 | B | | 86.87 (1.89) | 67.62 (4.55) | 77.25 (3.08) | 81.34 (2.57) |
| | | √ | 87.47 (2.64) | 70.49 (4.04) | 78.98 (3.24) | 82.44 (3.25) |
| | ScS | | 87.24 (1.95) | 73.43 (4.11) | 80.34 (2.94) | 82.78 (2.56) |
| | | √ | 87.97 (1.64) | 75.01 (2.63) | 81.49 (2.12) | 83.76 (2.04) |
| | W | | 85.82 (1.52) | 73.68 (2.89) | 79.75 (1.99) | 81.60 (1.81) |
| | | √ | 86.12 (1.05) | 75.66 (1.85) | 80.89 (1.30) | 82.34 (1.18) |
| | Ensemble | | 87.76 (1.31) | 76.67 (2.31) | 82.22 (1.76) | 83.95 (1.63) |
| | | √ | 87.99 (1.87) | 78.14 (3.01) | 83.06 (2.43) | 84.50 (2.30) |

Values are five-fold mean percentages with sample standard deviations in percentage points in parentheses. CMCL indicates cross-modality contrastive learning; B, diffuse blue-light illumination; ScS, sclerotic-scatter illumination; W, diffuse white-light illumination. A check mark indicates CMCL inclusion. The PVT patient-level rows reproduce baseline-voting and full-framework results; component-specific attribution requires identical aggregation and other training choices. See the configuration-comparison Methods. PVT indicates Pyramid Vision Transformer. Single-modality rows do not involve patient-level cross-modality aggregation. The patient-level rows without CMCL use hard modality voting, whereas the patient-level rows with CMCL use learned feature fusion.

**Table 9. Configurations differing in metadata, limbus cropping, and contrastive regularization.**

| Metadata Usage | Limbus Cropping | Contrastive Regularization | F1 Score | | | Accuracy |
|---|---|---|---|---|---|---|
| | | | Fungal | Bacterial | Macro-F1 | |
| √ | | | 87.24 (0.89) | 74.67 (2.25) | 80.96 (1.50) | 83.04 (1.22) |
| | √ | √ | 86.68 (1.68) | 75.97 (3.18) | 81.32 (2.43) | 82.86 (2.20) |
| √ | √ | | 88.02 (1.45) | 77.02 (2.68) | 82.52 (2.06) | 84.26 (1.88) |
| √ | √ | √ | 87.99 (1.87) | 78.14 (3.01) | 83.06 (2.43) | 84.50 (2.30) |

Values are five-fold mean percentages with sample standard deviations in percentage points in parentheses. A check mark indicates component inclusion. All four configurations used the MLP-based feature-fusion model. Rows 1 and 3 provide the matched comparison for limbus cropping, rows 2 and 4 provide the matched comparison for metadata, and rows 3 and 4 provide the matched comparison for the contrastive regularization.

**Table 10. Paired patient-level comparisons between TPMKA and baseline models.**

| Comparator | Accuracy, comparator / TPMKA (%) | Difference (pp) | McNemar P | Holm-adjusted P | Bacterial F1 difference, pp (95% CI) |
|---|---|---|---|---|---|
| ResNet-50 | 82.86 / 84.50 | +1.64 | 0.0260 | 0.1162 | +2.82 (0.89 to 4.80) |
| DenseNet-121 | 82.67 / 84.50 | +1.83 | 0.0232 | 0.1162 | +2.75 (0.57 to 4.88) |
| SwinV2-T | 83.53 / 84.50 | +0.97 | 0.1881 | 0.5643 | +1.89 (-0.08 to 3.82) |
| DeepIK | 83.47 / 84.50 | +1.03 | 0.1957 | 0.5643 | +1.63 (-0.47 to 3.72) |
| PVT_v2_b1 | 83.95 / 84.50 | +0.55 | 0.4847 | 0.5643 | +1.47 (-0.52 to 3.45) |

Accuracy and bacterial-F1 effect estimates are differences between arithmetic means of the five outer-fold metrics. Accuracy was compared using two-sided exact McNemar tests on paired out-of-fold correctness, with Holm correction across the five comparisons. Bacterial-F1 confidence intervals are unadjusted percentile intervals from 10,000 paired patient resamples within fixed outer-fold, country, and class strata. The same sampled patients were used for both models. CI indicates confidence interval; pp, percentage points.

**Table 11. Patient-level bootstrap confidence intervals for the six primary models.**

| Model | Population | Macro-F1, % (95% CI) | Accuracy, % (95% CI) | AUROC, % (95% CI) |
|---|---|---|---|---|
| ResNet-50 | Overall | 81.09 (79.65–82.50) | 82.86 (81.52–84.13) | 88.49 (87.13–89.81) |
| ResNet-50 | India | 63.33 (59.55–66.76) | 82.68 (81.06–84.22) | 79.42 (76.13–82.56) |
| ResNet-50 | US | 63.60 (57.26–69.04) | 83.31 (80.98–85.63) | 68.13 (61.11–74.87) |
| DenseNet-121 | Overall | 81.01 (79.46–82.54) | 82.67 (81.28–84.07) | 88.30 (86.96–89.55) |
| DenseNet-121 | India | 65.74 (61.94–69.22) | 82.85 (81.14–84.56) | 78.86 (75.63–82.00) |
| DenseNet-121 | US | 62.61 (56.79–67.81) | 82.25 (79.71–84.78) | 65.46 (58.36–72.46) |
| SwinV2-T | Overall | 81.82 (80.35–83.22) | 83.53 (82.25–84.80) | 89.14 (87.87–90.38) |
| SwinV2-T | India | 66.48 (62.55–70.04) | 84.13 (82.51–85.67) | 81.37 (78.19–84.39) |
| SwinV2-T | US | 60.65 (55.12–65.55) | 82.04 (79.72–84.37) | 68.36 (61.78–74.89) |
| DeepIK | Overall | 81.87 (80.28–83.45) | 83.47 (82.07–84.92) | 89.05 (87.73–90.35) |
| DeepIK | India | 68.73 (64.99–72.22) | 83.96 (82.17–85.67) | 81.60 (78.48–84.53) |
| DeepIK | US | 64.94 (58.94–70.35) | 82.24 (79.49–84.79) | 68.72 (61.84–75.35) |
| PVT_v2_b1 | Overall | 82.22 (80.77–83.66) | 83.95 (82.61–85.23) | 89.77 (88.50–91.02) |
| PVT_v2_b1 | India | 65.77 (61.89–69.41) | 83.87 (82.25–85.49) | 82.11 (78.97–85.08) |
| PVT_v2_b1 | US | 67.38 (61.13–72.71) | 84.15 (81.61–86.50) | 70.54 (64.09–76.89) |
| TPMKA | Overall | 83.06 (81.58–84.56) | 84.50 (83.16–85.84) | 89.93 (88.71–91.11) |
| TPMKA | India | 69.35 (65.59–72.87) | 84.56 (82.86–86.26) | 83.92 (81.02–86.66) |
| TPMKA | US | 65.98 (59.75–71.38) | 84.36 (82.04–86.68) | 66.96 (59.93–73.77) |

Each cell gives the arithmetic mean of five held-out-fold metrics and its percentile 95% confidence interval. Intervals were obtained from 10,000 patient-level resamples within fixed outer-fold, country, and organism-class strata. Intervals describe sampling uncertainty conditional on the fitted models and observed group composition. AUROC indicates area under the receiver operating characteristic curve; CI, confidence interval. US indicates United States.

**Table 12. Patient and image distribution in the Original and Balanced datasets.**

| Dataset | Country | Class | Patients, n | % of dataset | % within country | B available | ScS available | W available | B excluded | ScS excluded | W excluded | B images | ScS images | W images | All images |
|---|---|---|---|---|---|---|---|---|---|---|---|---|---|---|---|
| Original | India | Fungal | 961 | 58.42 | 82.00 | 954 | 956 | 959 | 7 | 5 | 2 | 3,546 | 3,644 | 3,653 | 10,843 |
| Original | India | Bacterial | 211 | 12.83 | 18.00 | 207 | 210 | 209 | 4 | 1 | 2 | 778 | 847 | 805 | 2,430 |
| Original | US | Fungal | 87 | 5.29 | 18.39 | 55 | 73 | 86 | 32 | 14 | 1 | 208 | 206 | 288 | 702 |
| Original | US | Bacterial | 386 | 23.47 | 81.61 | 247 | 294 | 382 | 139 | 92 | 4 | 930 | 925 | 1,328 | 3,183 |
| Original | All | All | 1,645 | 100.00 | 100.00 | 1463 | 1533 | 1636 | 182 | 112 | 9 | 5,462 | 5,622 | 6,074 | 17,158 |
| Balanced | India | Fungal | 87 | 25.00 | 50.00 | 87 | 87 | 87 | 0 | 0 | 0 | 310 | 336 | 330 | 976 |
| Balanced | India | Bacterial | 87 | 25.00 | 50.00 | 87 | 87 | 87 | 0 | 0 | 0 | 321 | 335 | 328 | 984 |
| Balanced | US | Fungal | 87 | 25.00 | 50.00 | 55 | 73 | 86 | 32 | 14 | 1 | 208 | 206 | 288 | 702 |
| Balanced | US | Bacterial | 87 | 25.00 | 50.00 | 87 | 87 | 87 | 0 | 0 | 0 | 367 | 310 | 317 | 994 |
| Balanced | All | All | 348 | 100.00 | 100.00 | 316 | 334 | 347 | 32 | 14 | 1 | 1,206 | 1,187 | 1,263 | 3,656 |

Percent of dataset uses the full dataset denominator; percent within country uses the country-specific denominator. The Balanced dataset is a patient-level subset with 87 patients in each country-class cell. B indicates blue-light illumination; ScS, sclerotic-scatter illumination; W, white-light illumination. The Balanced analysis changes both sample size and country-class composition and therefore does not isolate the effect of balancing. US indicates United States. Available indicates patients with at least one image from the corresponding illumination modality. Excluded indicates patients without that modality who were omitted from the corresponding modality-specific analysis but remained eligible for patient-level fusion through a zero-vector representation. In the Balanced dataset, the India-fungal, India-bacterial, and US-bacterial groups were sampled from patients with all three modalities. All 87 US-fungal patients were retained, including 32 without B images, 14 without ScS images, and 1 without W images.

**Table 13. Fungal and bacterial sensitivities and pooled out-of-fold confusion counts for TPMKA.**

| Setting | Population | Fungal sensitivity, % (95% CI) | Bacterial sensitivity, % (95% CI) | True F → F | True F → B | True B → F | True B → B |
|---|---|---|---|---|---|---|---|
| Original | Overall | 89.22 (87.69–90.74) | 76.21 (73.70–78.89) | 935 | 113 | 142 | 455 |
| Original | India | 94.59 (93.14–95.95) | 38.88 (32.24–45.54) | 909 | 52 | 129 | 82 |
| Original | US | 30.13 (20.85–39.48) | 96.63 (94.81–98.19) | 26 | 61 | 13 | 373 |
| Resampling | Overall | 87.60 (85.88–89.32) | 76.72 (73.87–79.58) | 918 | 130 | 139 | 458 |
| Resampling | India | 91.78 (90.02–93.45) | 50.75 (44.13–57.29) | 882 | 79 | 104 | 107 |
| Resampling | US | 41.63 (31.57–51.11) | 90.93 (88.07–93.77) | 36 | 51 | 35 | 351 |
| Group DRO | Overall | 86.64 (84.73–88.45) | 78.55 (75.87–81.23) | 908 | 140 | 128 | 469 |
| Group DRO | India | 91.36 (89.59–93.13) | 48.39 (42.19–55.02) | 878 | 83 | 109 | 102 |
| Group DRO | US | 34.51 (24.38–44.77) | 95.07 (92.74–97.15) | 30 | 57 | 19 | 367 |
| Resampling + Group DRO | Overall | 86.17 (84.26–87.98) | 75.88 (72.87–78.89) | 903 | 145 | 144 | 453 |
| Resampling + Group DRO | India | 90.64 (88.77–92.41) | 49.36 (42.74–55.96) | 871 | 90 | 107 | 104 |
| Resampling + Group DRO | US | 37.06 (27.71–46.54) | 90.42 (87.56–93.26) | 32 | 55 | 37 | 349 |

Sensitivities are arithmetic means of the five held-out outer-fold values; 95% confidence intervals were obtained from 10,000 patient-level resamples within fixed outer-fold, country, and organism-class strata. Confusion counts pool all out-of-fold patients, with rows representing the reference class and columns the predicted class. Consequently, a pooled count ratio may differ slightly from the fold-mean sensitivity. F indicates fungal; B, bacterial; CI, confidence interval; Group DRO, group distributionally robust optimization; US, United States.

## Supplementary figures

**Figure 4. Patient-level receiver operating characteristic (ROC) curves in each held-out fold for the six primary models. Each fold is plotted separately; summary values are arithmetic means and sample standard deviations of the five fold AUROCs. Bacterial keratitis is the positive class. The diagonal dashed line indicates chance-level discrimination. AUROC indicates area under the receiver operating characteristic curve; SD, standard deviation. ResNet-50 and DenseNet-121 are residual and densely connected convolutional-network comparators, respectively; SwinV2-T is the tiny Swin Transformer V2 comparator; PVT_v2_b1 denotes the b1 variant of Pyramid Vision Transformer version 2.**

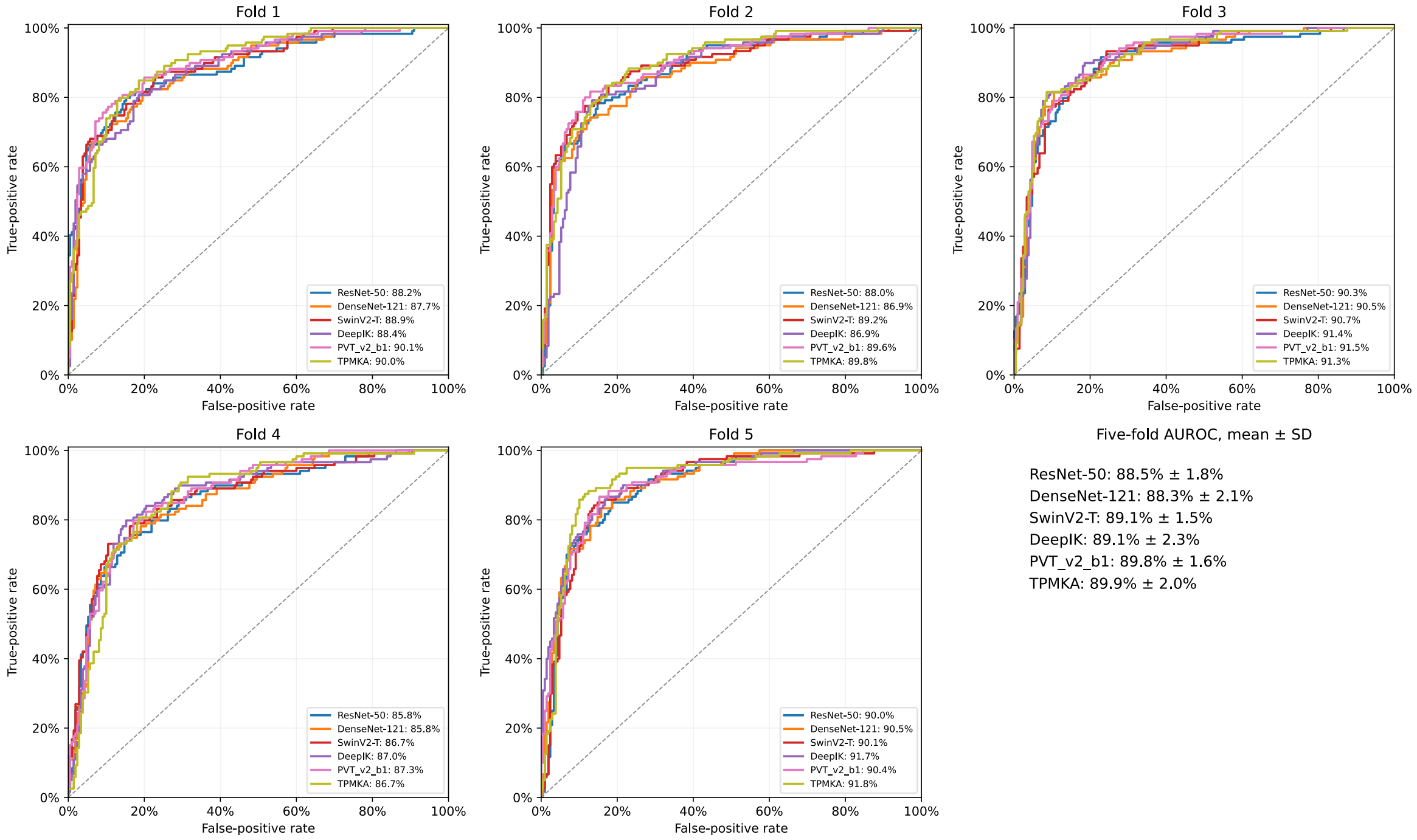